\newcommand\hi{\hbox{H{\scriptsize I}}}

\newcommand\oiii{\hbox{[O{\scriptsize III}]}}
\newcommand\oii{\hbox{[O{\scriptsize II}]}}
\newcommand\sii{\hbox{[S{\scriptsize II}]}}
\newcommand\lya{Lyman $\alpha$}

\newcommand\halpha{H$\alpha$}

\documentclass[iop,revtex4]{emulateapj}
\usepackage{color,lscape,amsmath}
\shorttitle{Indirect Evidence for Escaping Ionizing Photons in Local Lyman Break Galaxy Analogs}
\shortauthors{Alexandroff et al.}

\begin{document}
\title{Indirect Evidence for Escaping Ionizing Photons in Local Lyman Break Galaxy Analogs} 
\author{Rachael M. Alexandroff, Timothy M. Heckman, Sanchayeeta Borthakur}
\affil{Center for Astrophysical Sciences, Department of Physics and Astronomy, Johns Hopkins University, Baltimore, MD 21218, USA.}
\email{rmalexan@jhu.edu}
\author{Roderik Overzier}
\affil{Observat$\acute{o}$rio Nacional, Ministry of Science, Technology and Innovation, Rio de Janeiro, Brazil.}	
\author{Claus Leitherer}
\affil{Space Telescope Science Institute, 3700 San Martin Drive, Baltimore, MD 21218, USA.}
\date{\today}

\label{firstpage}

\begin{abstract}
A population of early star-forming galaxies is the leading candidate for the re-ionization of the universe.  It is still unclear, however, what conditions and physical processes would enable a significant fraction of the ionizing (Lyman continuum) photons to escape from these gas-rich galaxies.  In this paper we present the results of the analysis of HST COS far-UV spectroscopy plus ancillary multi-waveband data of a sample of 22 low-redshift galaxies that are good analogs to typical star-forming galaxies at high-redshift.  We measure three parameters that provide indirect evidence of the escape of ionizing radiation (leakiness): (1) the residual intensity in the cores of saturated interstellar low-ionization absorption lines, which indicates incomplete covering by that gas in the galaxy. (2) The relative amount of blue-shifted \lya~line emission, which can indicate the existence of holes in the neutral hydrogen on the front-side of the galaxy outflow, and (3) the relative weakness of the [SII] optical emission lines that trace matter-bounded HII regions.  We show that our residual intensity measures are only negligibly affected by infilling from resonance emission lines. We find all three diagnostics agree well with one-another.  We use these diagnostics to rank-order our sample in terms of likely leakiness, noting that a direct measure of escaping Lyman continuum has recently been made for one of the leakiest members of our sample. We then examine the correlations between our ranking and other proposed diagnostics of leakiness. We find a good correlation with the equivalent width of the \lya~emission line, but no significant correlations with either the flux ratio of the [OIII]/[OII] emission lines or the ratio of star-formation rates derived from the (dust-corrected) far-UV and  \halpha~luminosities.  Turning to galaxy properties, we find the strongest correlations with leakiness are with the compactness of the star-forming region (SFR/area) and the speed of the galactic outflow.  This suggests that extreme feedback- a high intensity of ionizing radiation and strong pressure from both radiation and a hot galactic wind- combines to create significant holes in the neutral gas.  These results not only shed new light on the physical mechanisms that can allow ionizing radiation to escape from intensely star-forming galaxies, they also provide indirect observational indicators that can be used at high-redshift where direct measurements of escaping Lyman continuum radiation are impossible.
\end{abstract}

\section{Introduction}
\label{sec:intro}
 
Current evidence from observations of the cosmic microwave background \citep[CMB;][]{Dunkley2009,Planck2013} and \lya~forest absorption in quasar spectra \citep[e.g.][]{Fan2006} suggests that the reionization of the universe occurred over an extended period between redshifts of $\sim$ 15 to 6.  Deep near-IR (NIR) imaging with the Hubble Space Telescope (HST) indicates that the UV luminosity density of early star-forming galaxies is high enough that they are the best candidates to provide the ionizing ultra-violet (UV) photons necessary for this process \citep[e.g.][]{Bouwens2012,Shull2012,Oesch2013,Robertson2015}.  For this to be possible, a mechanism is required by which Lyman continuum photons, produced by young stars in the centers of gas-rich star forming galaxies, can escape into the intergalactic medium (IGM) and the escape fraction of Lyman continuum photons from these galaxies must be relatively high: $f_{esc} \gtrsim 0.2$ \citep[e.g.][]{Bouwens2011,Robertson2013}.

Studies of high redshift galaxies have yet to uncover a large sample of galaxies with the required high escape fraction of ionizing photons.  For example, \citet{Iwata2009} only find evidence of escaping continuum photons in 17 of 198 high redshift ($z \simeq 3.1$) Lyman Break galaxies (LBGs) and \lya~emitters (LAEs), \citet{Vanzella2010,Vanzella2012} find evidence for escaping Lyman continuum photons in only one galaxy of 102 LBGs studied at a redshift of $z \sim 4$ and \citet{Nestor2013} detect Lyman continuum in 6 of 41 LBGs and 17 of 91 LAEs at $z \gtrsim 3.0$.  Even when escaping Lyman continuum photons are detected, in most high redshift studies $f_{esc}$ is usually only around $5\%$ \citep[e.g.][and references above]{Steidel2001,Giallongo2002,Fernandez2003,Inoue2005,Shapley2006,Mostardi2013}.  This is higher than the upper limit \citep[e.g.][]{Leitherer1995,Heckman2001,Deharveng2001,Grimes2009} or value \citep[e.g.][]{Leitet2011,Leitet2013} of a few percent measured in most local galaxies or our own Milky Way \citep{Putman2003} as well as the value measured in galaxies at redshift $z \sim 1$ \citep[e.g.][]{Malkan2003,Siana2007,Cowie2009,Siana2010}.  Yet, it is still not high enough to reionize the universe.  Such a small number of detections is perhaps not surprising as an \hi~column density of only $1.4 \times 10^{17} $cm$^{-2}$ at the Lyman edge is enough to produce unit optical depth and most star-forming regions from galactic disks to local starburst and high redshift star forming galaxies are gas-rich with column densities of $10^{21}-10^{24}$ cm$^{-2}$ \citep{Kennicutt1998,Genzel2010}.  

One possible solution is to assume that galaxies have a ``patchy" coverage of high column density, optically thick neutral gas clouds located in an optically-thin neutral medium.  Such a ``picket fence" model would allow ionizing photons to escape along optically thin lines of sight not blocked by an optically thick neutral cloud.  In such a model, the escape fraction, $f_{esc}$, of ionizing radiation is simply the percent of the galaxy not covered by neutral clouds, $f_c$ ($f_{esc} = 1- f_c$).  However, the conditions required to produce such a geometry may be relatively unusual \citep[e.g.][]{Dove2000,Clarke2002,Fujita2003,Wise2009,Razoumov2010}. 

Conducting such studies at high redshift (z $>$ 4) is inherently difficult. At the epoch of reionization, the spectral region shortward of \lya~is opaque making it impossible to study. Even for galaxies at a redshift above $z \simeq$ 4 the distribution of intergalactic neutral hydrogen causes an attenuation of $> 1.8$ mag below \lya~\citep{Inoue2014}.  In addition, studies of high redshift star-forming galaxies, with the exception of small samples of lensed galaxies, suffer from low spectral resolution and poor signal/noise for individual objects.  A sample of low redshift galaxies similar to high redshift star forming galaxies would allow us to search directly for escaping Lyman continuum photons with high sensitivity, to calibrate/validate indirect indicators of escaping ionizing photons that could be employed at high redshift, and provide clues as to the physical conditions and mechanisms that foster the escape of ionizing radiation. 
 
A previous study \citep[][hereafter H11]{Heckman2011} used HST Cycle 17 data from the Cosmic Origins Spectrograph (COS) in the far-UV (FUV; HST Program 11727: PI T. Heckman) to study 8 starbursting local galaxies known as Lyman Break Analaogs (LBAs).  Due to the relatively poor sensitivity of COS beyond the Lyman edge for objects with $z\sim0.1-0.25$, these observations were made at longer wavelengths and the escape fraction of ionizing radiation was inferred from several indirect indicators.  Previously, \citet[][hereafter O09]{Overzier2009} and H11 compiled a list of indirect indicators that may provide evidence for the presence of escaping ionizing continuum photons.  These include: 1) a large ratio of the star-formation rate (SFR) derived from dust-corrected UV to that derived from dust-corrected \halpha~possibly due to the loss of ionizing photons.  2) weak [\ion{S}{2}]~emission relative to \halpha~or [\ion{N}{2}], possibly indicating matter-bounded HII regions.  3) low-ionization interstellar absorption lines that were optically-thick but with significant residual intensity in their cores, indicative of ``holes" in that gas. And 4) \lya~emission lines with a significant ratio of blue-shifted:red-shifted emission again indicative of holes in the neutral ISM that would permit leakage of both \lya~emission and ionizing continuum radiation.  

H11 also found that objects with a dominant central object (DCO; O09)-- a massive ($M_* \gtrsim 10^9M_{\bigodot}$) and very compact (R$\sim$100pc) starburst-- were more likely to show strong supernovae-driven winds ($\sim$1000 kms/s) and ``patchy" neutral hydrogen coverage with a range of dust-free escape fractions from $0.3 \leq f_{esc}^{dust free} \leq 0.6$.  However, this sample was not large enough for a full statistical analysis.  Thus, in HST Cycle 20 we successfully proposed for COS FUV-spectroscopy of an additional 14 LBAs (HST Program 13017 : PI T. Heckman).  In this paper we report the results of an analysis of these new data and re-analysis of the previous Cycle 17 data which allows us to draw statistically robust conclusions about the degree to which these different indirect indicators of escaping ionizing radiation agree with one another and ascertain which properties of the starburst correlate best with these indicators.

Our results are made even more relevant by recent results from \citet{Borthakur2014} that show direct evidence for escaping Lyman continuum from one of the LBAs in H11.  The object in question, SDSSJ092159.38+450912.3 (J0921) contains a DCO with a relatively high ($\sim25\%$) predicted escape fraction of Lyman continuum photons from the variety of indirect methods listed above.  Follow-up observations using COS below the Lyman edge found a luminosity, $\lambda L_{\lambda}$ at a rest wavelength of $910 \rm \AA$ of $5.0 \times 10^{42}$ erg s$^{-1}$, implying a dust-free escape fraction of $f_{esc}^{dust free} = 21 \pm 5\%$, similar to that predicted  by indirect indicators.  

The rest of the paper proceeds as follows; Section \S~\ref{sec:ss} describes our sample selection, \S~\ref{sec:data} and \S~\ref{sec:ancillary} describe our observations and data analysis.  Section \S~\ref{sec:results} compares the indirect indicators of escaping Lyman continuum photons with each other and \S~\ref{sec:discussion} examines their dependence on the galaxy/starburst properties.  We offer our conclusions in Section \S~\ref{sec:conclusion}.  The names of all objects in the paper are derived from the hour and minutes of their J2000 right ascension coordinates as taken from the SDSS. Full coordinates are provided in Table 1. Throughout we adopt a flat $\Lambda$CDM cosmology using $\Omega_{\Lambda}=0.7$, $\Omega_M=0.3$,and $H_0=70~$km s$^{-1}$ Mpc$^{-1}$ \citep{Spergel2007}.  

\section{Sample Selection}
\label{sec:ss}
As described above, the existence of a local population of LBA galaxies has been established over the past decade \citep{Heckman2005}. This sample was originally selected by combining optical data from the Sloan Digital Sky Survey \citep[SDSS;][]{York2000} with UV data from the Galaxy Evolution Explorer \citep[GALEX;][]{Martin2005} to match the UV properties of LBGs using the sample criteria:

\begin{enumerate}
  \item No detectable AGN contribution to continuum emission
  \item FUV luminosity of $L_{FUV} \geq 2 \times 10^{10}L_{\bigodot}$ measured at $1550 \rm \AA$
  \item FUV effective surface brightness (mean FUV surface brightness within the half-light radius of the SDSS u-band) of $I_{FUV} \geq 10^9L_{\bigodot}$ kpc$^{-2}$
\end{enumerate}

Subsequent research has shown that this LBA sample is matched with LBGs in terms of morphology, size, UV luminosity, SFR, mass, gas fraction, velocity dispersion, metallicity, and dust content \citep{Hoopes2007,Basu-Zych2007,Overzier2008,Basu-Zych2009,Overzier2009,Goncalves2010,Overzier2010,Overzier2011,Goncalves2014}.  Thus, we are able to use the LBA sample to look for clues as to how outflows of gas and ionizing radiation from early star-forming galaxies enriched and reionized the IGM.  A list of all the LBAs observed in our two HST campaigns with COS along with details of the observations and their properties is provided in table \ref{tab:lba}.  Objects were chosen for follow-up with COS that had the highest FUV continuum fluxes (as evidenced by their GALEX photometry) and appeared to be the most compact (this applies only for the sample from H11 where previous UV imaging from HST was available) so as to provide the best possible spectral resolution.  The sample presented here is based on LBAs selected in the original sample of \citet{Heckman2005}, supplemented with new targets found in the latest crossmatch (Overzier et al. in prep.) between SDSS data release 7 \citep[DR7;][]{Abazajian2009} and GALEX GR6.

\section{HST Observations and Data Analysis}
\label{sec:data}

\begin{figure*}
\includegraphics[scale=0.5,trim=170 400 150 75,clip=true]{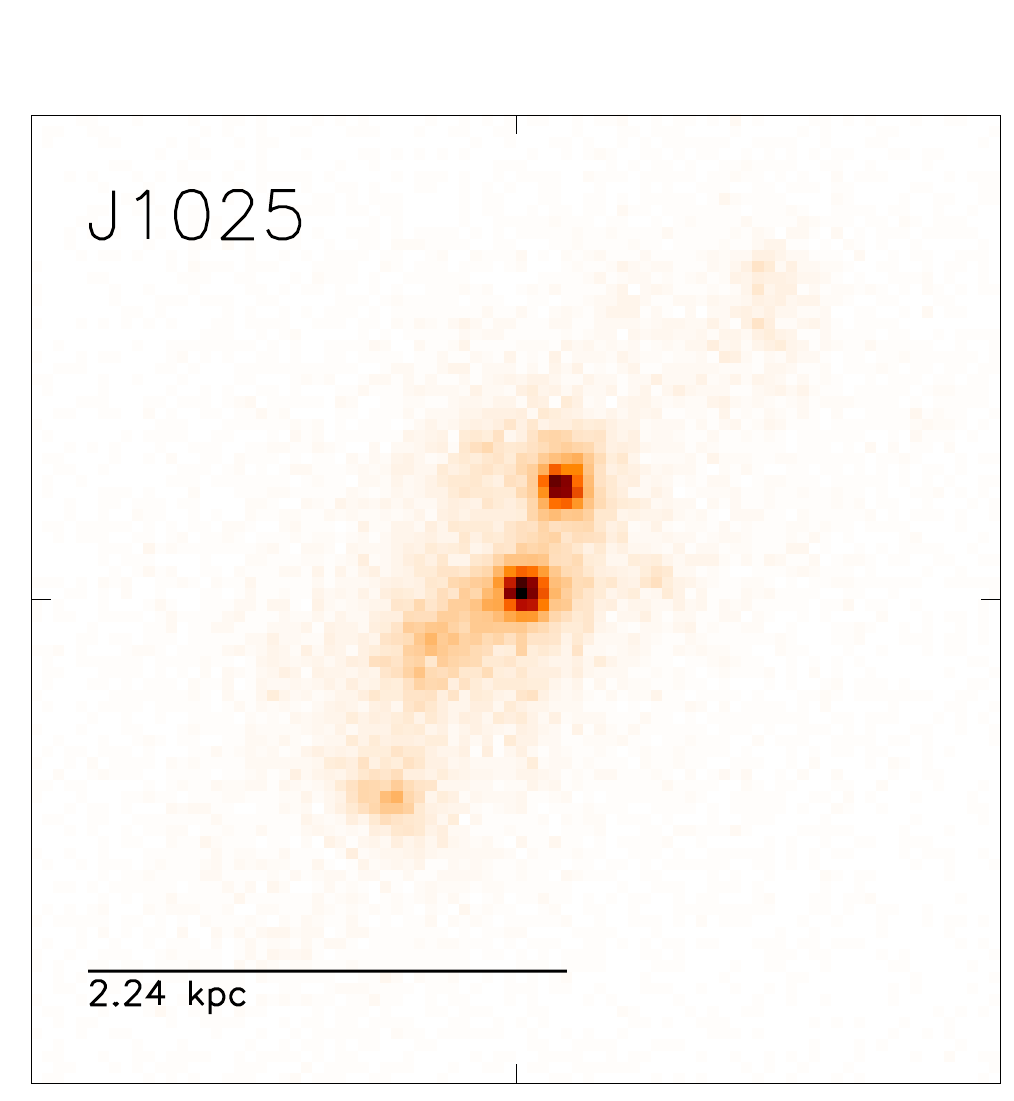}
\includegraphics[scale=0.5,trim=150 400 150 75,clip=true]{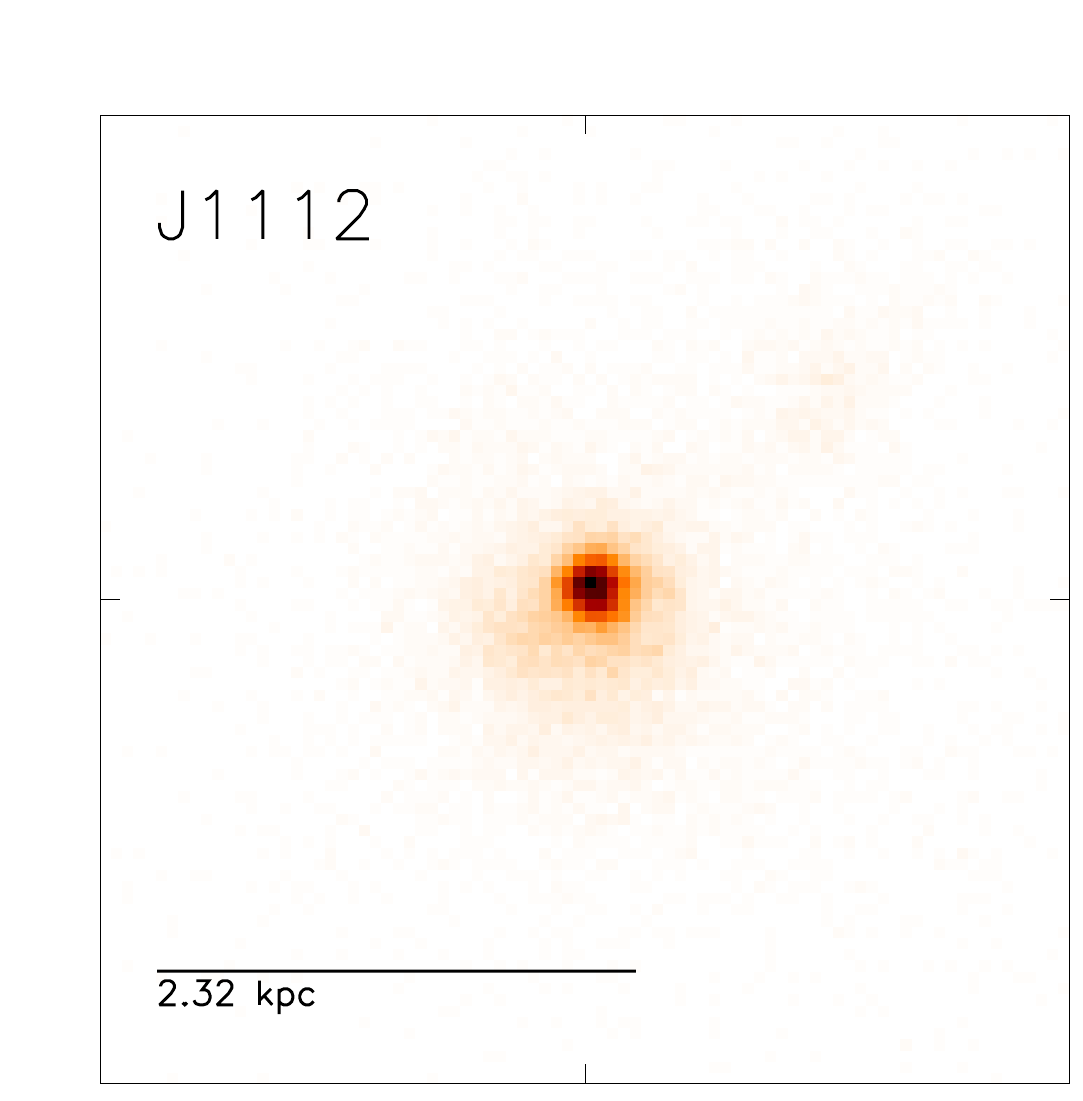}
\includegraphics[scale=0.5,trim=170 400 150 75,clip=true]{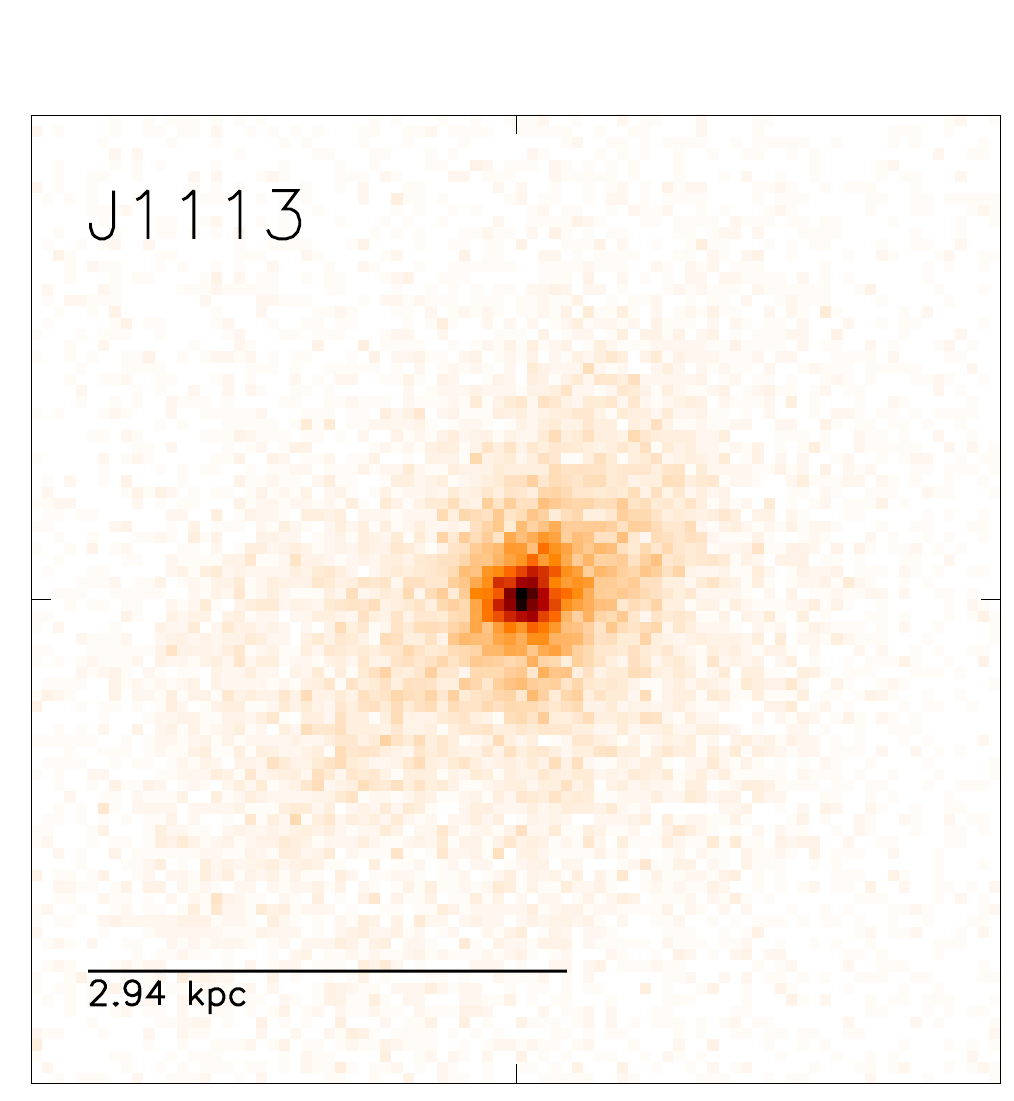}
\includegraphics[scale=0.5,trim=170 400 150 75,clip=true]{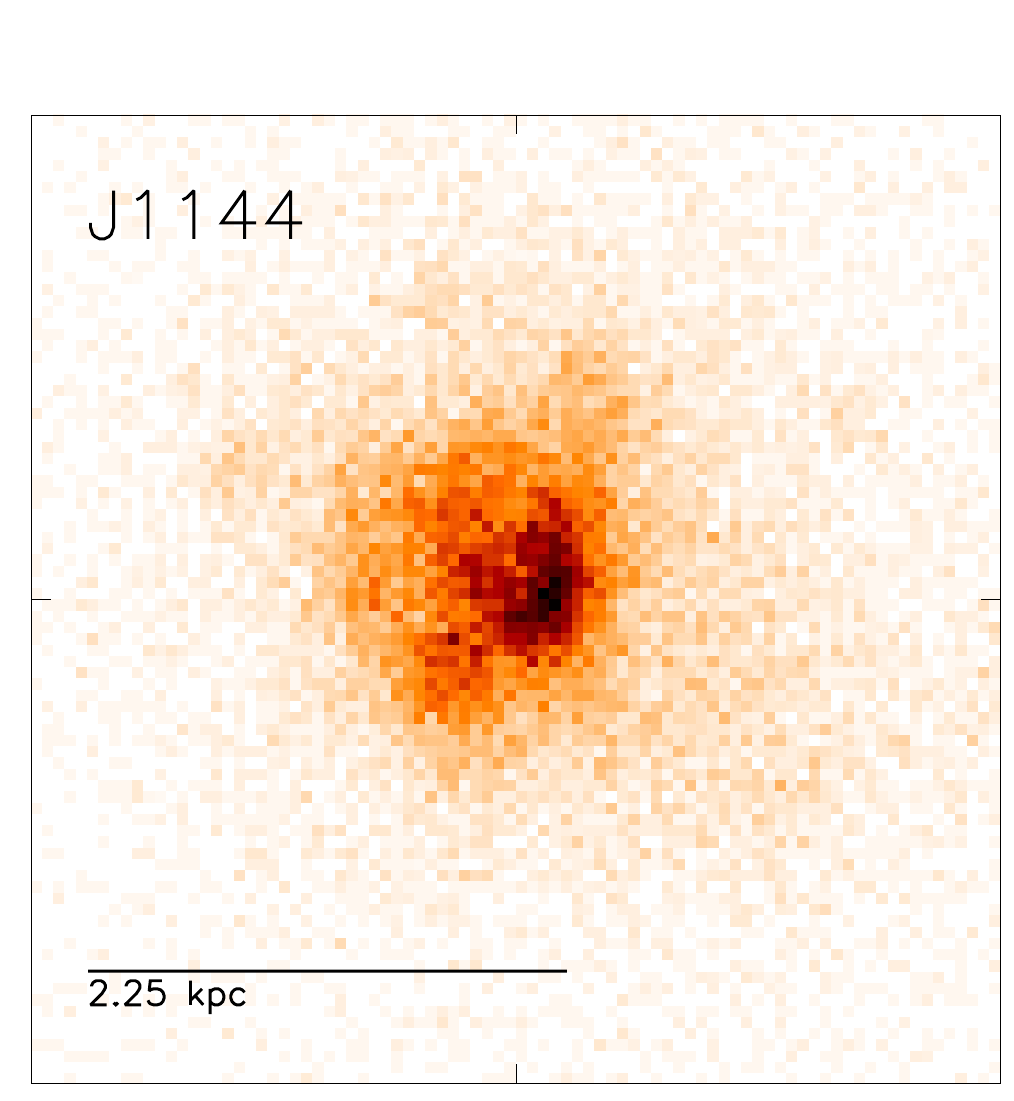}
\includegraphics[scale=0.5,trim=170 400 150 75,clip=true]{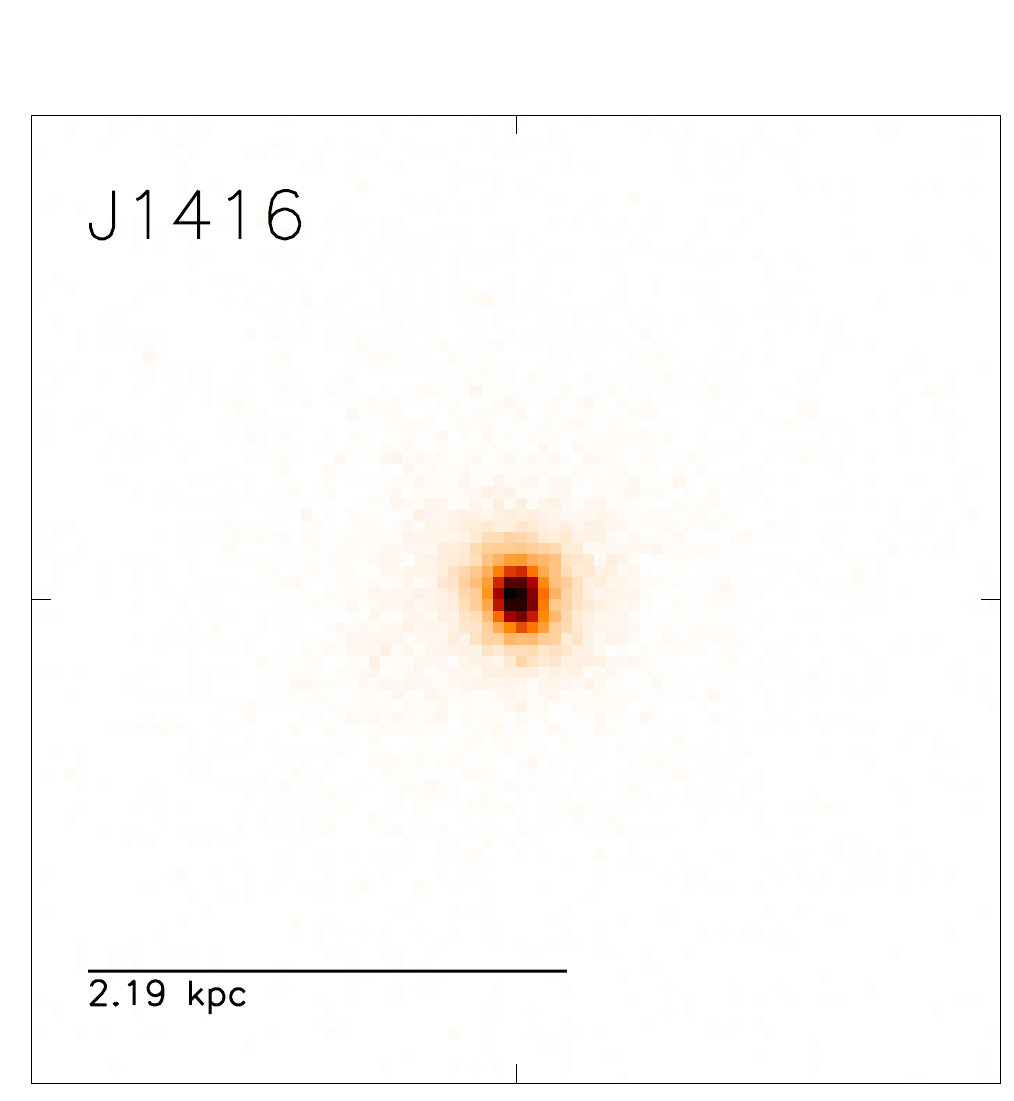}
\includegraphics[scale=0.5,trim=170 400 150 75,clip=true]{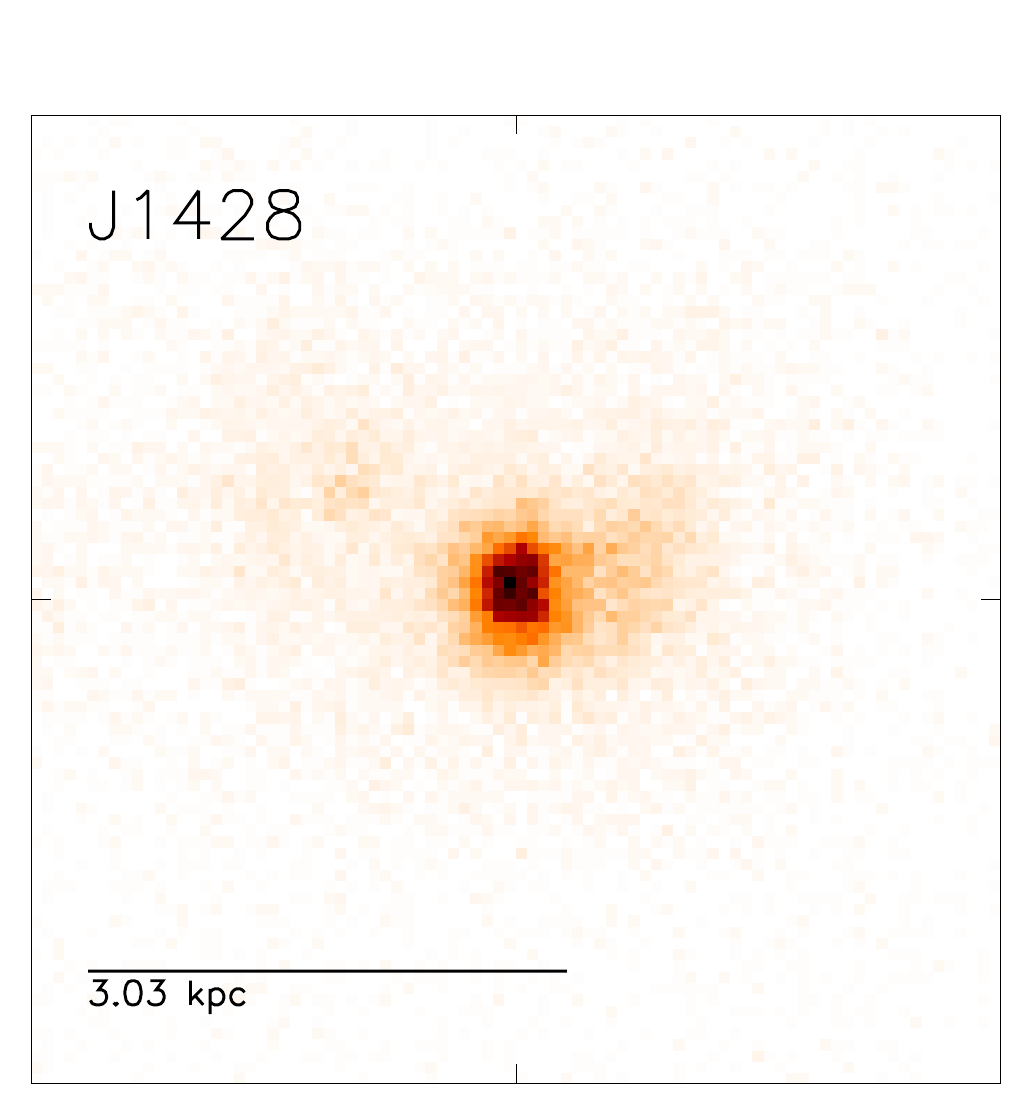}
\includegraphics[scale=0.5,trim=170 400 150 75,clip=true]{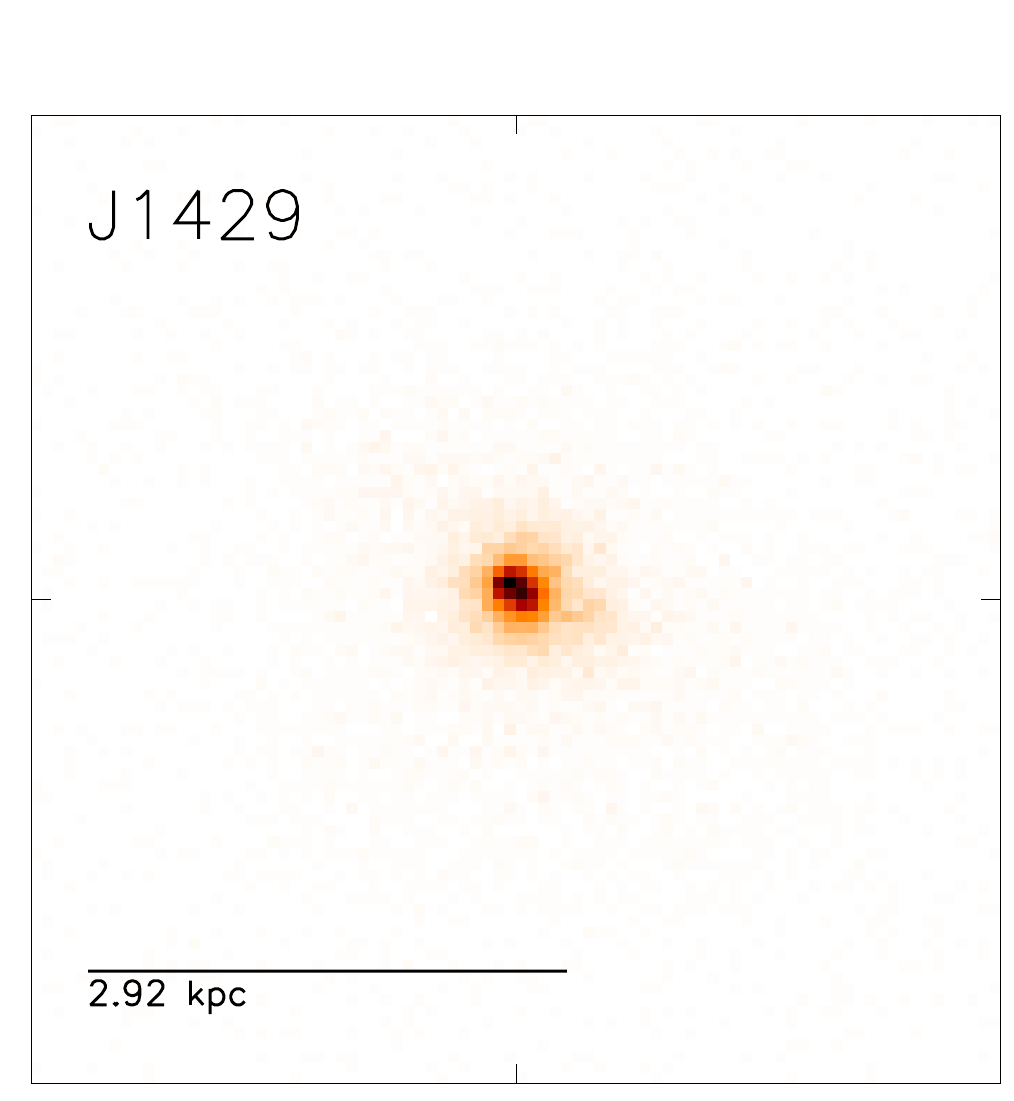}
\includegraphics[scale=0.5,trim=100 400 150 75,clip=true]{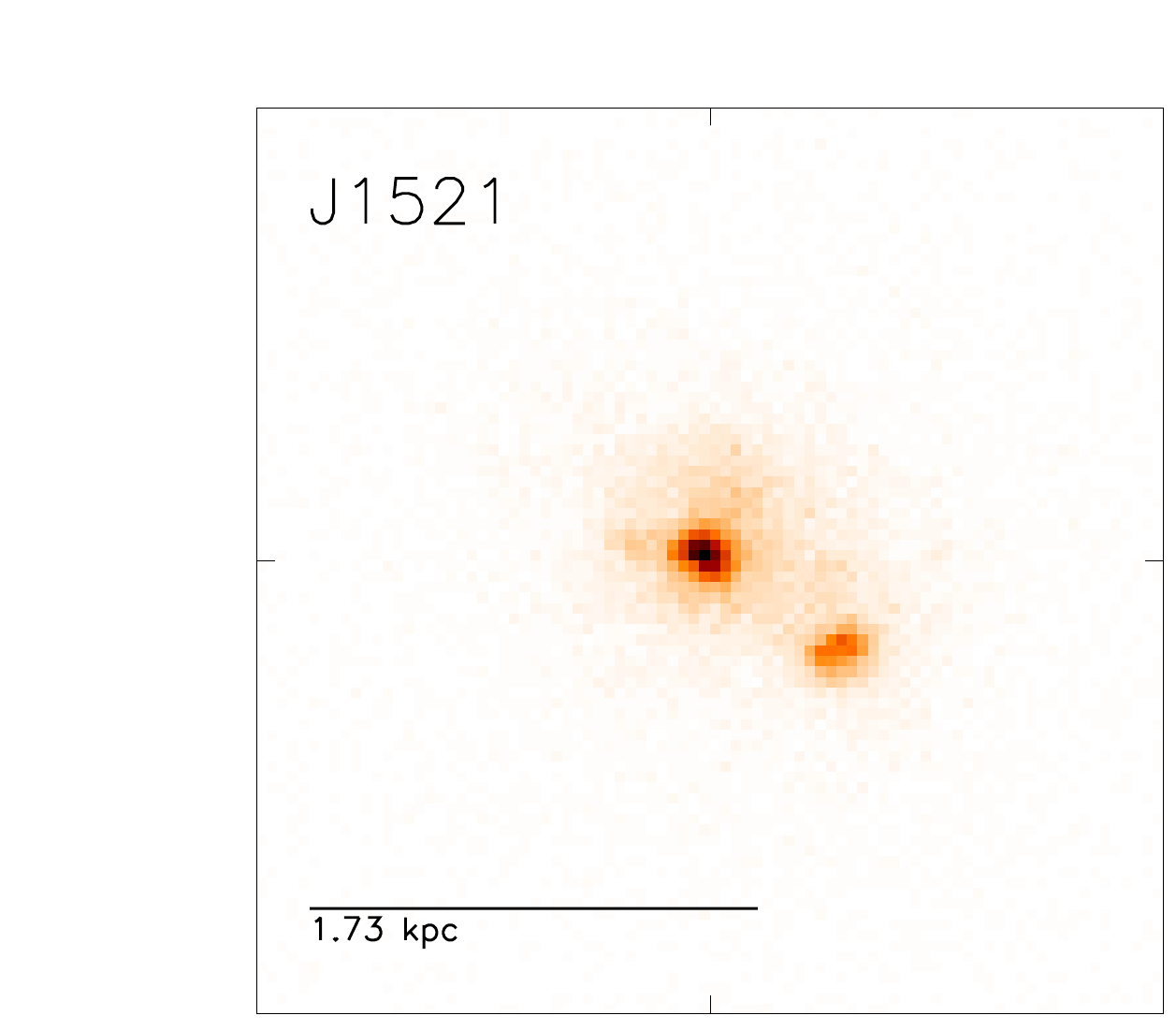}
\includegraphics[scale=0.5,trim=100 400 150 75,clip=true]{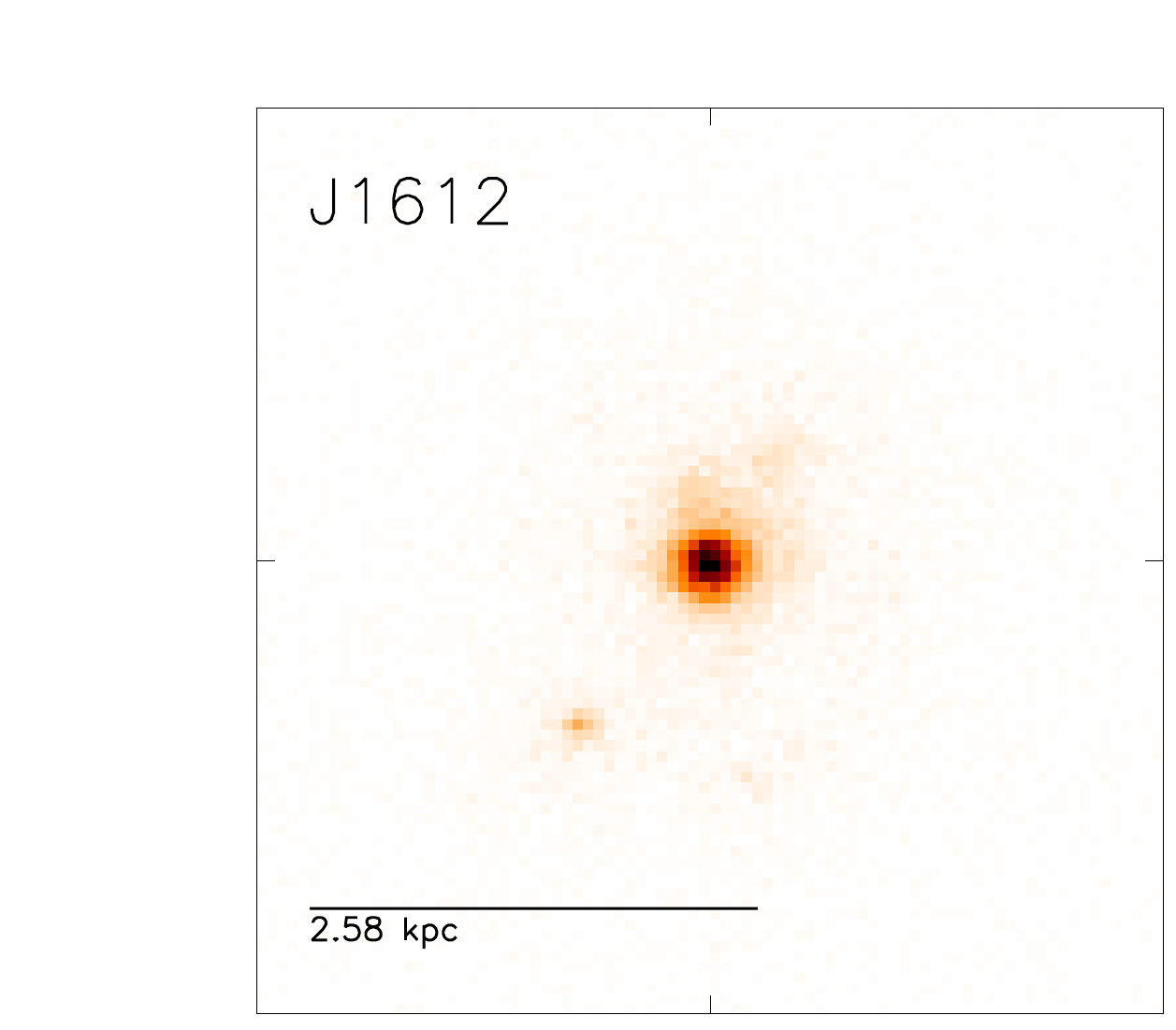}
\caption{\small NUV images of the new LBA sample.  These images were taken using COS in 100 sec exposures primarily for acquisition purposes.  Each image is 2'' per side with the corresponding physical scale for 1'' marked.  The objects were imaged with COS in the MIRRORA configuration.}
\label{fig:postage1}
\end{figure*}

\begin{figure*}
\includegraphics[scale=0.5,trim= 170 400 150 75,clip=true]{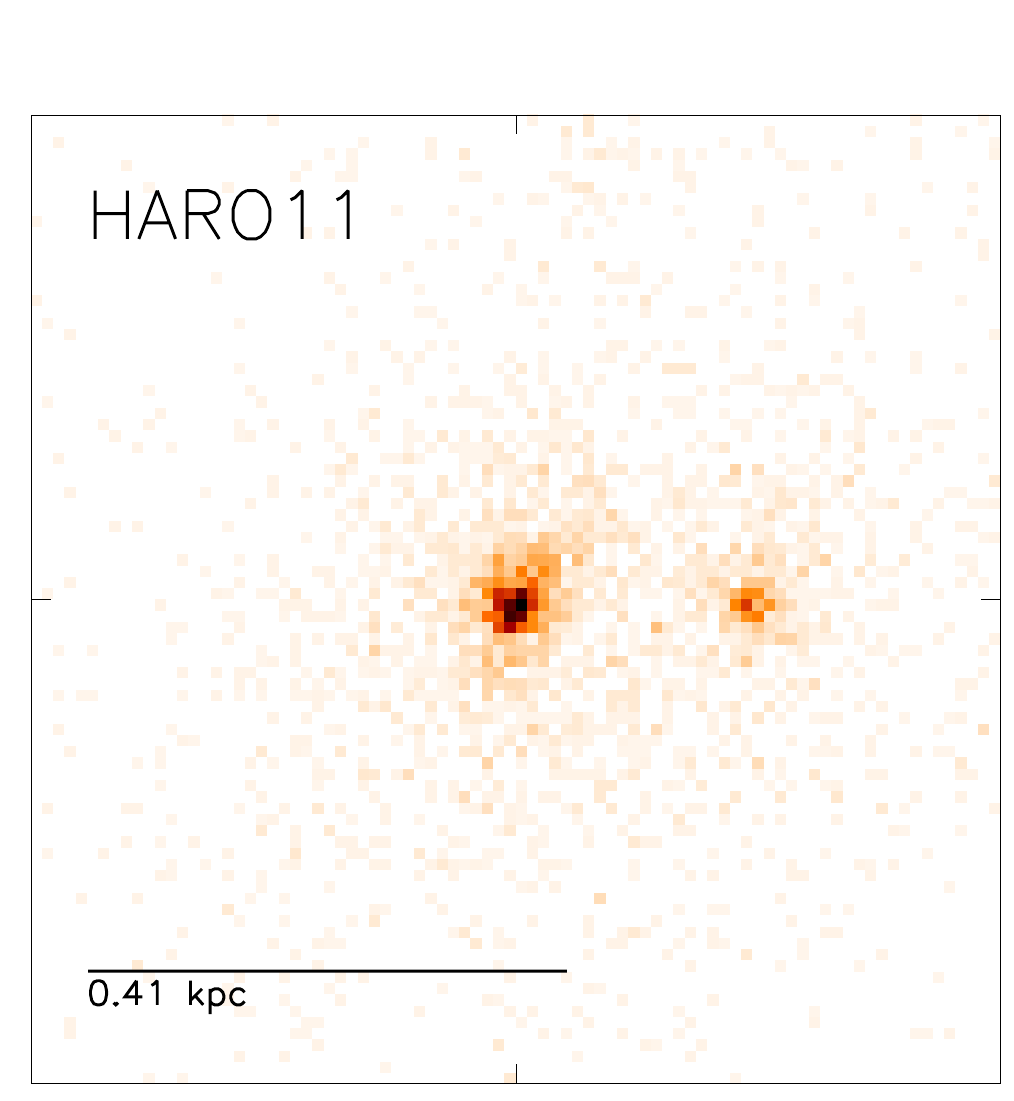}
\includegraphics[scale=0.5,trim=170 400 150 75,clip=true]{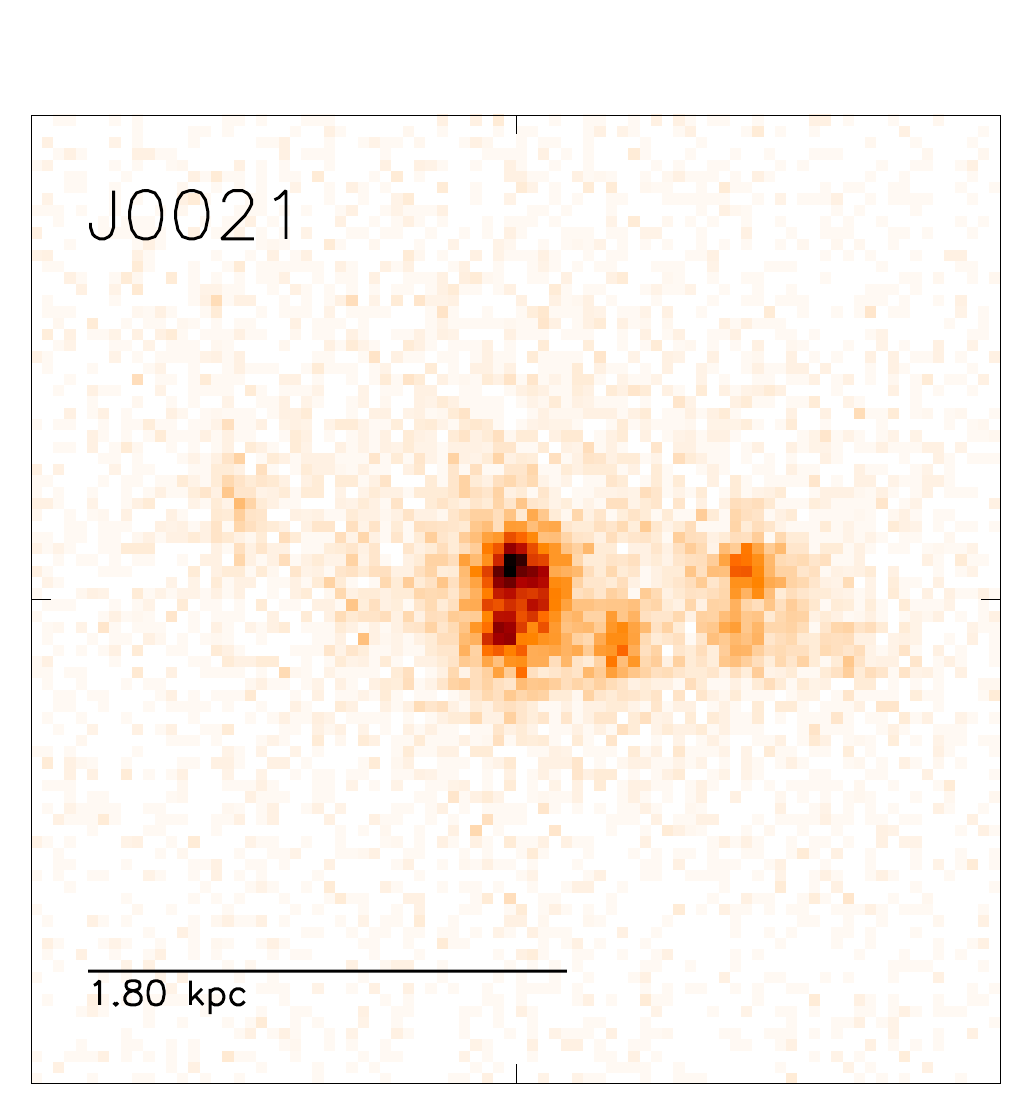}
\includegraphics[scale=0.5,trim=170 400 150 75,clip=true]{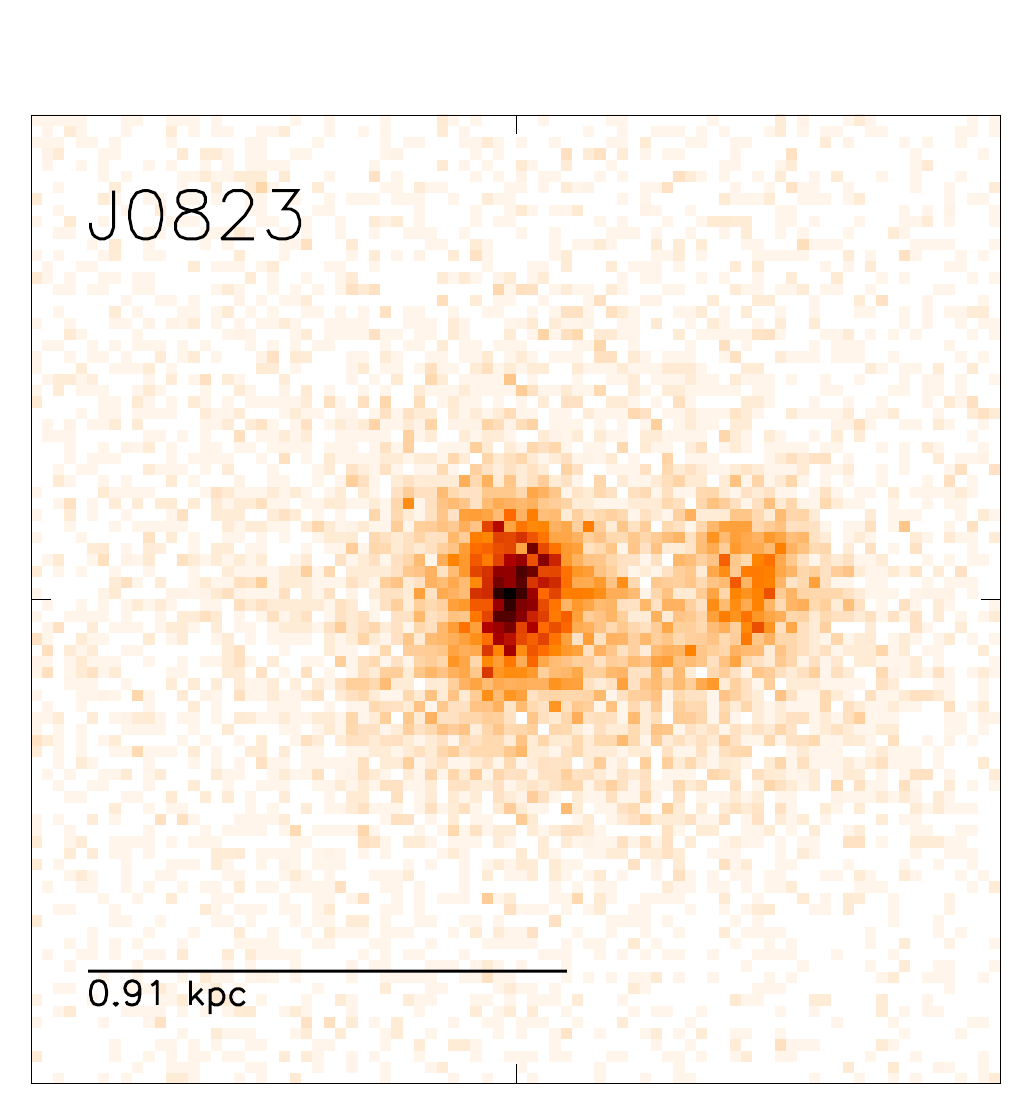}
\includegraphics[scale=0.5,trim=170 400 150 75,clip=true]{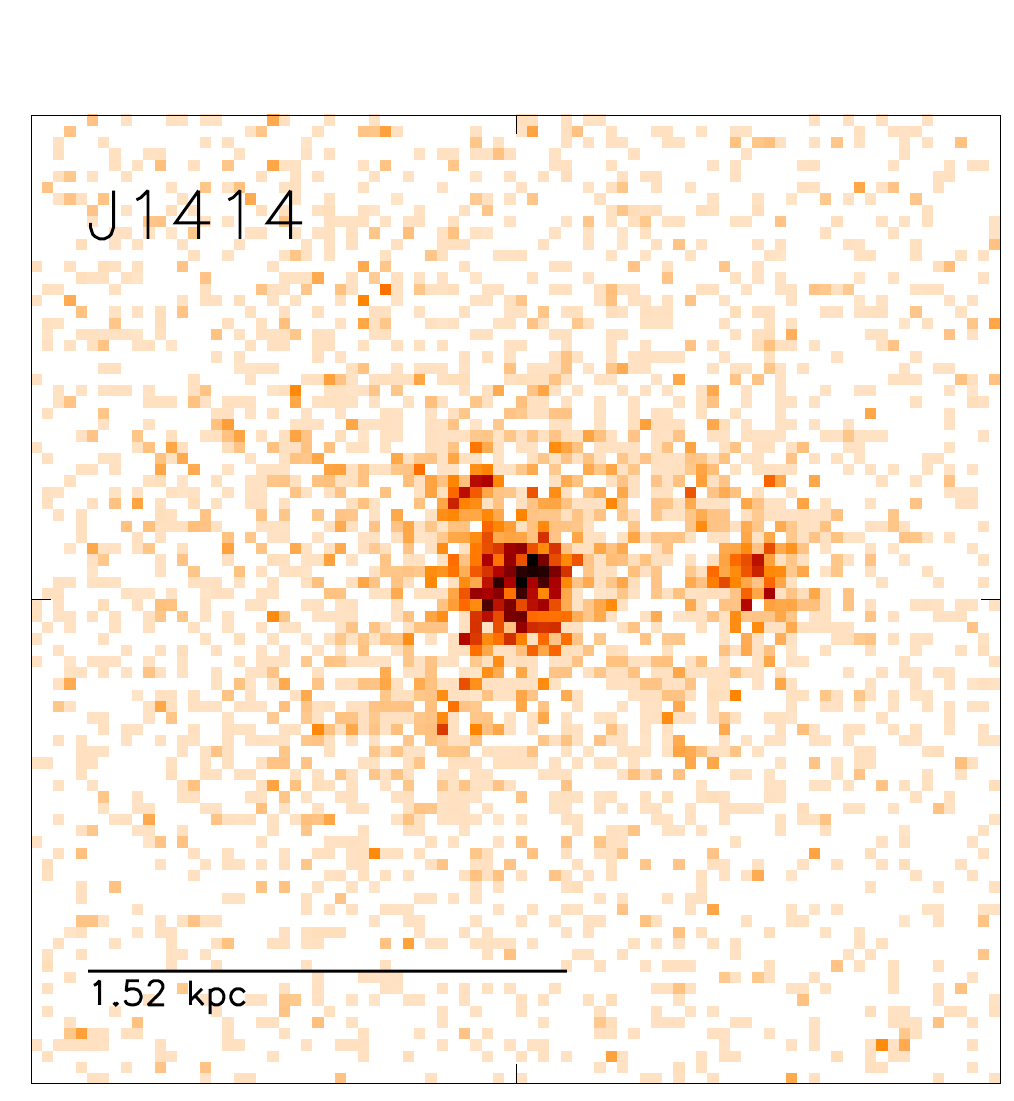}
\includegraphics[scale=0.5,trim=100 400 150 75,clip=true]{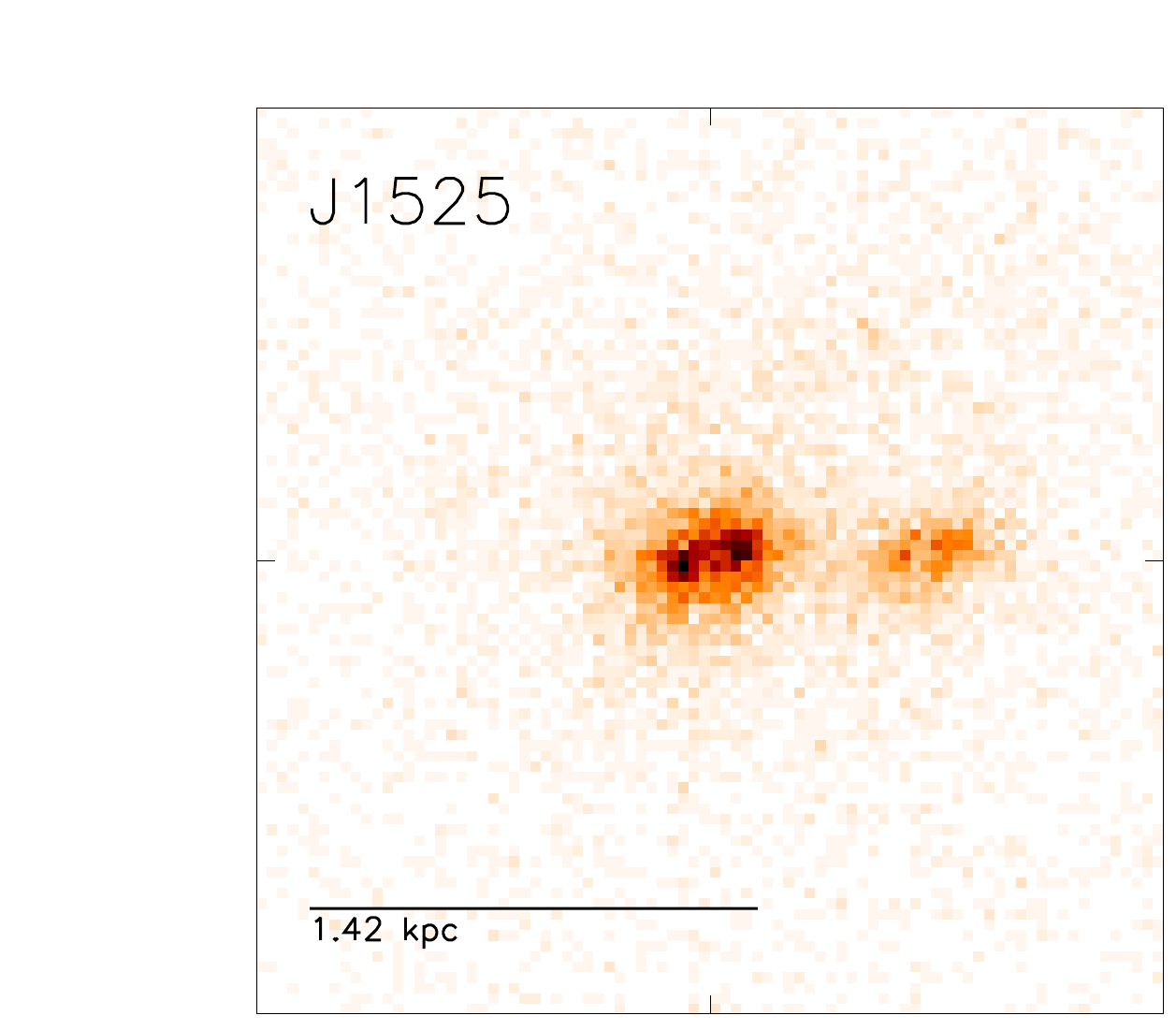}
\caption{\small NUV images of the new LBA sample.  These images were taken using COS in 100 sec exposures primarily for acquisition purposes.  Each image is 2'' per side with the corresponding physical scale for 1'' marked.  The objects were imaged with COS in the MIRRORB configuration. MIRROR B produces a secondary image seen here to the right of the primary image for each object.}
\label{fig:postage2}
\end{figure*}

Our observations were conducted using the COS instrument aboard HST in cycle 20.  Each observation includes a 100~sec acquisition image in the near-UV(NUV) as well as a longer spectroscopic exposure (1-2 orbits depending on the FUV flux expected to pass through the COS aperture).  Our spectral observations use the COS FUV G130M and G160M medium resolution gratings (R$\sim$17000).  This spectral resolution ensures that the spectral line-spread function is smaller than the velocity dispersion of the ISM within the LBA and so allows us to resolve important absorption- and emission line features.  Our observations span the observed wavelength range from $\lambda = 1160 \rm \AA - 1780 \rm \AA$~and are designed to allow maximum spectral coverage from rest-frame \ion{O}{6}~($\lambda~1035\rm \AA$) which probes coronal gas at T$\sim10^5-10^6$ K to \ion{C}{4}~($\lambda~1550\rm \AA$) and so probes the O star population through stellar winds.  In addition, this wavelength range covers several important ISM transitions such as \lya, the low-ionization transitions of \ion{Si}{2}~($\lambda~1190,1193,1260,1304,1526\rm \AA$) and \ion{C}{2}~($\lambda~1334\rm \AA$) and the intermediate ionization transition of \ion{Si}{3}~($\lambda~1206\rm \AA$).  Of the newest sample six objects are at $z \leq 0.11$ meaning the \ion{O}{6}~line is out of reach while three are at $z \geq 0.16$ meaning \ion{C}{4}~is not covered.  Postage stamp images of the new sample of 14 LBAs in our HST COS Cycle 20 sample are shown in Figures \ref{fig:postage1} and \ref{fig:postage2}.  Note that the images shown are the acquisition images and that those in figure \ref{fig:postage2} were taken with the MIRROR B setting and so show a secondary image on the detector.  UV images of the previous H11 sample can be found in O09. 

\subsection{COS Data}
\begin{figure*}
\includegraphics[angle = 90,scale=0.8,trim=50 80 80 80,clip=true]{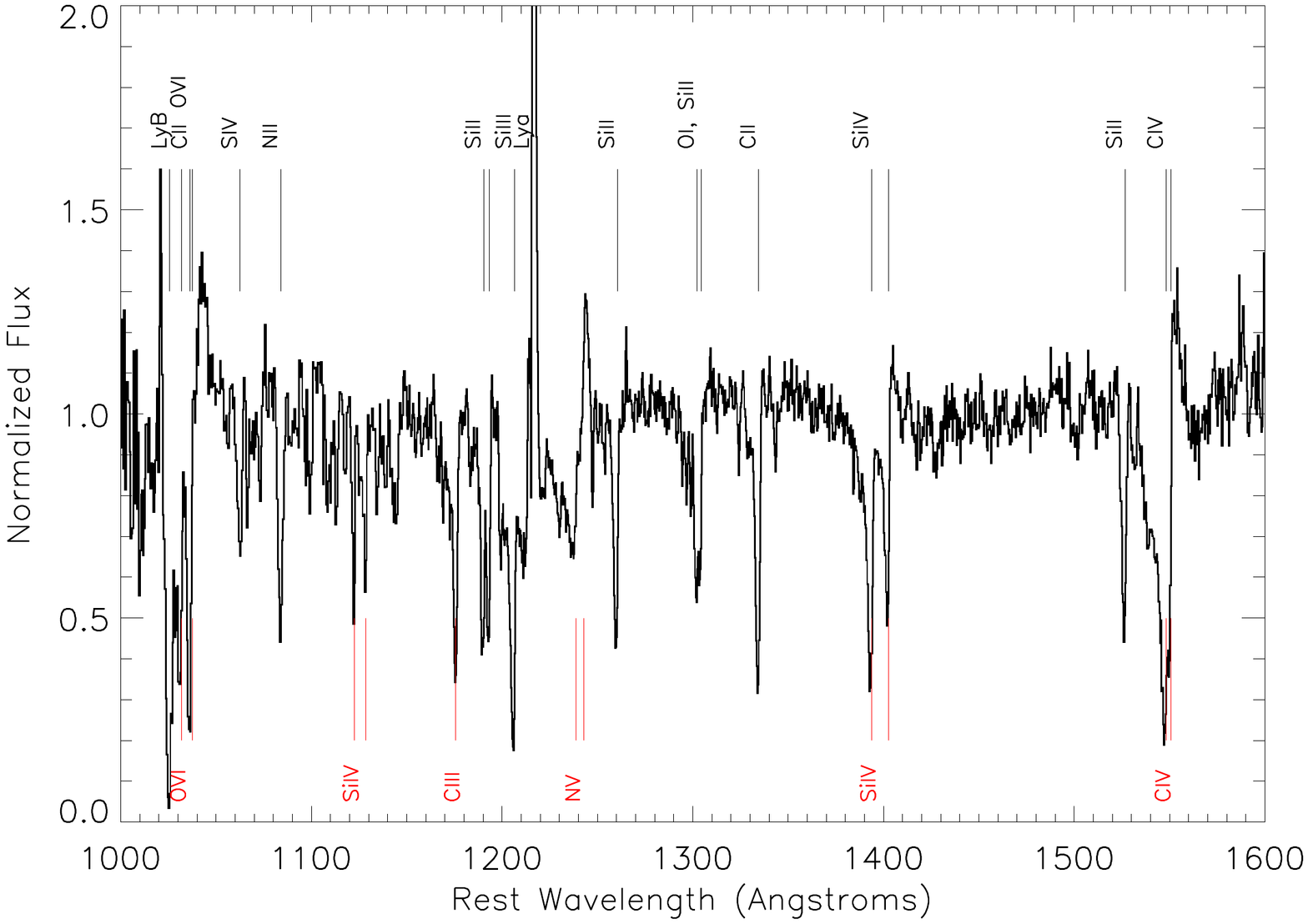}
\caption{Coadd of our full sample of 22 COS spectra.  No weighting of individual spectra was applied. All spectra have been moved to the same wavelength grid as the SB99 models.  Important ISM lines are noted above in black and important stellar features are noted below in red.  Lines with both stellar and interstellar components are indicated in both colors.}
\label{fig:coadd}
\end{figure*}

All spectra were run through a suite of procedures similar to those described in H11.  Data were first retrieved from the MAST archive and were processed through the standard COS pipeline.  Then, the G130M and G160M spectra were merged and rebinned by 21 pixels (corresponding to a spectral bin size of $\sim 50$km s$^{-1}$); no attempt was made to smooth the data at this point.  The data were then normalized by fitting a power law followed by a low-order polynomial function to the continuum as defined by \citet{Calzetti1994} and \citet[][defined in regions of the spectra designed to avoid strong emission and absorption features both in the observed and rest frame]{Leitherer2002} and then dividing the spectra by this polynomial. The full sample was combined to make a single high signal-to-noise spectrum shown in figure \ref{fig:coadd}.     

Our goal in this paper is to measure the properties of the interstellar gas. To that end, we undertook a procedure to remove the spectral features produced by stars (e.g. photospheric and stellar wind lines).
Starburst99 \citep[from now on SB99;][]{Leitherer1999} was used to generate synthetic spectra based on stellar evolutionary synthesis models.  We produced models based on two star formation histories: an instantaneous starburst with no subsequent star formation and a continuous starburst with a constant rate of star formation.  The stellar population was parameterized by a Kroupa initial mass function \citep[IMF;][]{kroupa2001}.  The metallicity of the stellar population was previously determined by O09 and we created two sets of model spectra with SB99 to bracket the observed metallicities.  The stellar population then evolves from the zero-age main sequence using the high mass loss evolutionary models of the Geneva Group.  The model spectrum we used are generated from model atmospheres, not an empirical library; these spectra are described in detail in \citet{Leitherer2010}.  We sample the model every two million years to generate the synthetic spectra for our model fitting.  Our observed COS data were interpolated to the wavelength grid of the spectra produced by our stellar evolutionary synthesis models. This caused some degradation in our spectral resolution, especially at shorter wavelengths.

\begin{figure*}
\includegraphics[scale=0.8,angle=90,trim=50 50 50 90,clip=true]{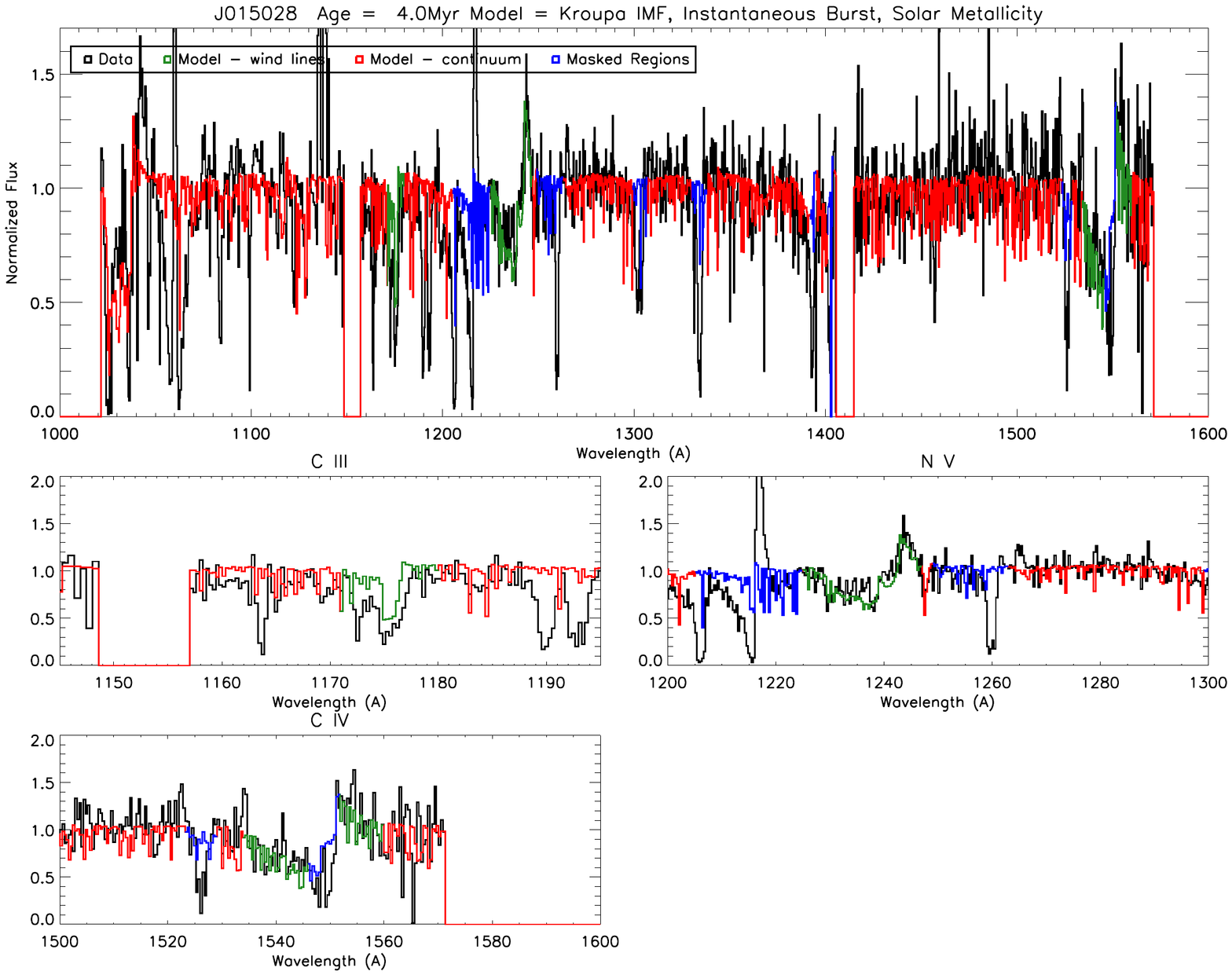}
\caption{\small Best fit model for a single LBA, J0150.  The data is shown in black and over plotted in colour is the best fit model originally generated by SB99.  The model has three components; in red is the model continuum (not used in the fit), in blue are regions of the model that are masked because they overlap with interstellar features (not used in the fit).  Finally, in green are the stellar wind lines we used for our chi-squared test.  This figured was generated using code from Dr. Christy Tremonti \citep{Tremonti2004}.}
\label{fig:model_fit}
\end{figure*}

The method we employed for finding the best fit model is also described in \citet{Tremonti2001}.  First, we do our own normalization of the SB99 models according to the same prescription used for the COS data.  We compare our data and models based only on a select few stellar wind lines generated by massive O stars.  For this analysis we used the regions surrounding \ion{C}{3}$\lambda~1175\rm \AA$ (1171 - 1180$\rm \rm \AA$),  \ion{N}{5}$\lambda~1239\rm \AA$(1225 - 1247$\rm \rm \AA$), and \ion{C}{4}$\lambda~1548\rm \AA$ (1534 -1560$\rm \rm \AA$), except for those objects for which \ion{C}{4}~is not available.  In two cases (J0823, J1113) a damped \lya~(DLA) system and circum-galactic medium of a nearby galaxy respectively obscured the \ion{N}{5}~feature, preventing us from using it in our comparison.  In one of these cases, J1113, there was also no \ion{C}{4}~available as the object was at a higher redshift and so the model comparison was tied to a single absorption feature.  There is a known problem with the \ion{O}{6} $\lambda~1038\rm \AA$ feature in SB99 due to the Doppler broadening of the Ly$\beta$ emission line generated by SB99 that prevents us from using it to fit our data.  However, we do find that our data shows rough consistency with the SB99 \ion{O}{6}~when this line is excluded from the fit.  We then calculated the chi-squared value over all selected regions for each model spectra and take the ``best-fit" model to be the model with the lowest chi-squared value.  We find that, in most cases, our LBAs are best fit by models with ages less than 10 million years whether instantaneous or continuous starbursts (see figure \ref{fig:model_fit}).  We then subtract the best fit model from our data to remove the stellar spectral component.  With the stellar continuum and line features removed, we are able to study the ISM signatures in detail.  

Unless otherwise stated all measurements can be assumed to have errors on the order of 10\%-15\% dominated by systematics in the polynomial fit to the continuum emission and subtraction of the SB99 models.

\subsection{Measurements of velocity outflows}
\label{ssec:velocity}

\begin{figure*}
\includegraphics[scale=0.8,angle=180]{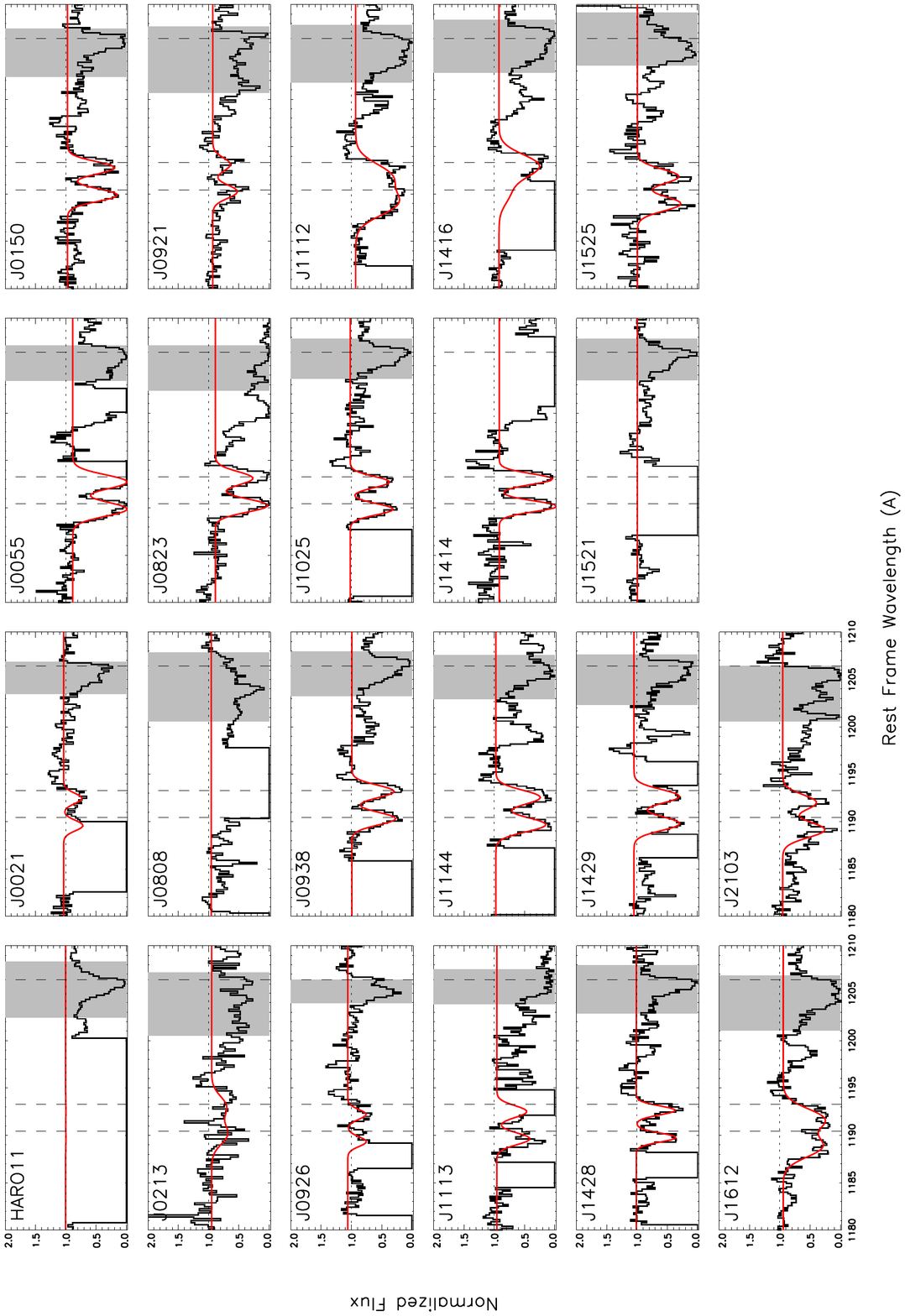}
\caption{\small Spectral region containing \ion{Si}{2}$\lambda~1190,1193\rm \AA$ and \ion{Si}{3}$\lambda~1206\rm \AA$ absorption features tracing the low- and mid-ionization ISM.  The spectra shown have been normalized and the best-fit stellar model subtracted so the normalized continuum flux is equal to 1.  Areas where MW features interfere with the spectra have been masked.  The rest wavelength for each line, determined by the SDSS redshift, is marked in black.  The gaussian model fit to both \ion{Si}{2}~transitions is shown in red.  Each absorption line is fit with a single gaussian component constrained such that both transitions as well as \ion{Si}{2}$\lambda$~1260$\rm \rm \AA$ have the same width and centroid but the flux in each transition is allowed to differ.  The region over which the outflow speed is calculated from \ion{Si}{3}~is indicated in grey. In the case of J1414, where SiIII was not available in the spectrum, the outflow velocity was calculated using the CII absorption feature. Notice the significant amount of residual flux in both \ion{Si}{2} lines as well as the asymmetric, blue profiles of all three lines.}
\label{fig:gaussian_model}
\end{figure*}

Strong outflows of gas are generic in intensely star-forming galaxies at low- and high-redshift. One signature of these outflows are the broad blueshifted interstellar absorption lines (see figure \ref{fig:gaussian_model}). H11 suggested that very high outflow speeds might be associated with galaxies with escaping Lyman continuum emission. In order to characterize the outflow speeds we use the SiIII $\lambda~1206\rm \AA$ transition as opposed to \ion{Si}{2} or \ion{C}{2} because it is typically the strongest interstellar absorption feature in the spectra of the LBAs allowing us to characterize outflow speeds even in cases where significant residual intensity makes other lines shallow.  In the case of J1414 where \ion{Si}{3} was not available in the spectrum, the outflow velocity was calculated using the \ion{C}{2} absorption feature.  We measure the flux-weighted line centroid of this feature, integrating over a spectral region selected by eye (shown in Figure \ref{fig:gaussian_model} in grey) to have detectable absorption. Future work (Heckman et al. 2015, in prep) will involve a more detailed analysis of the outflow velocities derived across the range of available ISM transitions.  We found a maximum outflow speed of 561 km s$^{-1}$ (J0808) with an average outflow speed of 375 km s$^{-1}$. Infilling of absorption lines from scattered emission could cause an offset between the measured and actual outflow velocities though this effect is likely small \citep[see section \ref{ssec:resonance} of this paper or][]{Scarlata2015, Prochaska2011}.

\subsection{Residual Flux in saturated ISM lines}
\label{ssec:res_flux}
H11 showed that the strong absorption lines tracing the low-ionization ISM in LBAs are saturated (optically thick) but do not always have zero residual intensity in the line center. They interpreted this as evidence for partial coverage of the FUV continuum (the starburst) by neutral gas \citep[as in e.g.][]{Heckman2000,Rupke2002,Quider2009,Quider2010}.  In this model, optically-thick neutral gas clouds are interspersed in an optically-thin medium and thus the percentage of the ionizing source not covered by neutral gas clouds is the same as the residual flux (percent) in saturated ISM lines.  This technique is increasingly used at higher redshift where permitted \citep[see, for example][]{Jones2013}. 

This is most clearly seen by examining the series of \ion{Si}{2}~transitions which span over an order of magnitude in oscillator strength (f): $\lambda~1190.416\rm \AA (f = 0.277)$, $\lambda~1193.290\rm \AA (0.575)$, $\lambda~1260.422\rm \AA (1.22)$, $\lambda~1304.370\rm \AA (0.093)$ $\lambda~1526.707\rm \AA (0.133)$\footnote{oscillator strengths from NIST Atomic Database: http://www.nist.gov/pml/data/asd.cfm}. 
If the UV sources were being viewed through a uniform covering of optically-thin clouds we would expect the residual intensity in each line to decrease as the oscillator strength (optical depth) increases. On the other hand, if our UV sources were only partially covered by optically-thick clouds, there would be no dependence of residual intensity on oscillator strength and instead the residual intensity at the core of each line ($I_0$) would be related to the covering factor ($f_c$) of neutral clouds ($I_0/I_{cont} = 1 - f_c$).  

We began by fitting the 1190$\rm \AA$, 1193$\rm \AA$, and 1260$\rm \AA$ lines, which should be the most optically thick. These absorption lines were fit with a Gaussian model using the IDL package MPFIT \citep{Markwardt2009}.  We constrained all three lines in a fit to have the same centroid and width in velocity space but allowed the line strength to vary so as to be able to study changes in the residual flux across different absorption lines. Residual flux was measured for each line by calculating the depth of the line at the centroid calculated from our model. We found that the residual flux in all three lines is largely consistent, and frequently large. 

\begin{figure}
\includegraphics[scale=0.5,trim = 60 350 0 100,clip = true]{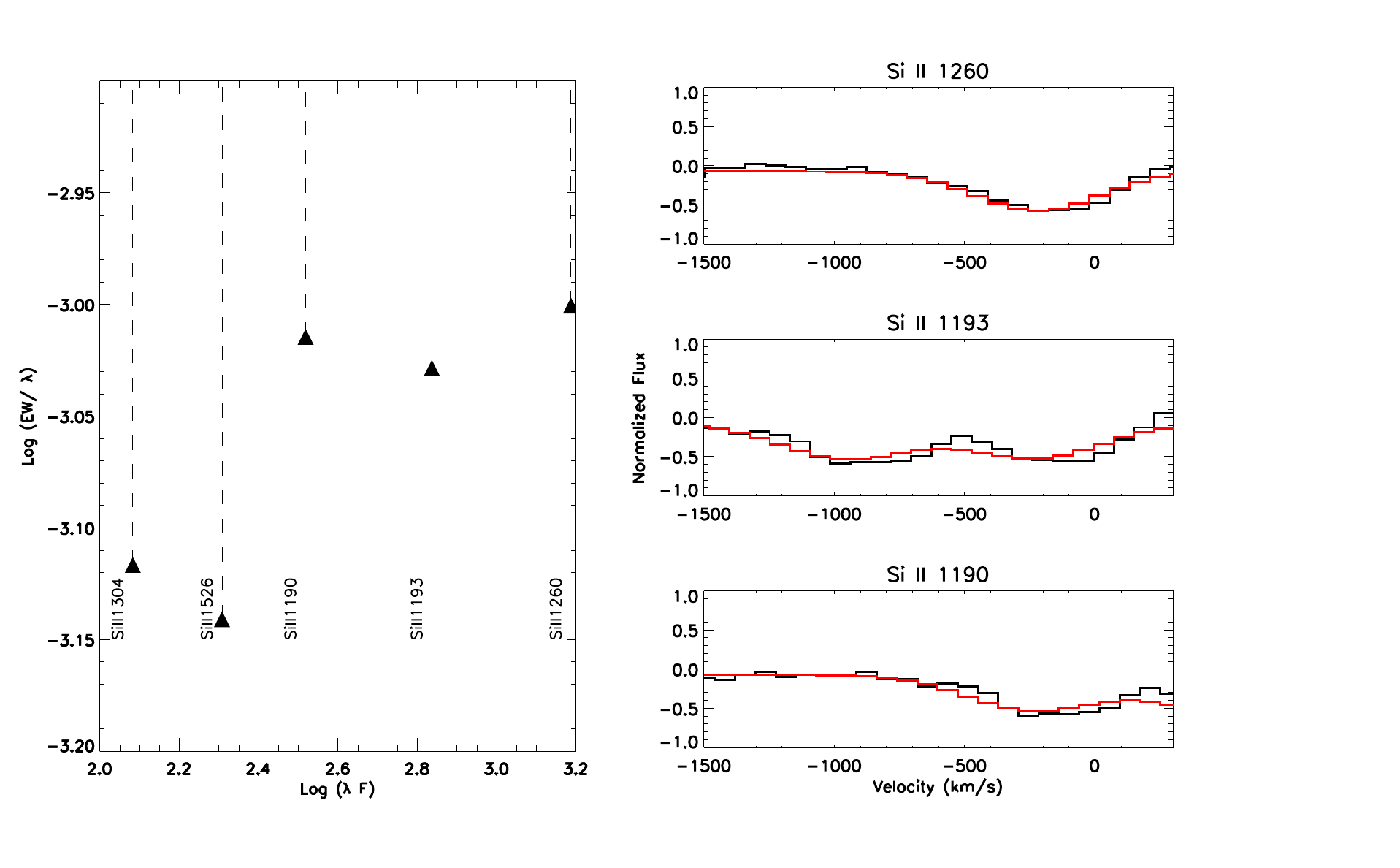}
\caption{\small Left: Evolution of \ion{Si}{2}~transition rest-frame equivalent width as a function of oscillator strength ($\log \lambda f$) also referred to as the ``curve of growth" for the co-added spectrum of our 22 LBAs.  We see the characteristic evolution from optically thin absorption ($EW/\lambda \propto f\lambda$) to optically thick absorption where the curve-of growth becomes flat.  This provides confidence that our strongest \ion{Si}{2}~absorption lines, particularly $\lambda~1260$ is indeed optically thick and any residual intensity in the core of the feature must be due to a less than 100\% covering of neutral gas.  Note that there is a transition of \ion{S}{3} at 1190.206$\rm \AA$ that contaminates the \ion{Si}{2} 1190.416$\rm \AA$ absorption line and probably explains why it unexpectedly appears strong than \ion{Si}{2}1193 in our figure. Right: Model fit to the three most prominent \ion{Si}{2}~transitions.  Our model uses a single gaussian for each absorption line constrained such that all lines have the same width and centroid, only the flux in each transition is allowed to differ.}
\label{fig:curve_of_growth}
\end{figure}

To further confirm that the strong \ion{Si}{2}~transitions in our spectrum were optically thick, we calculated the evolution in equivalent width (EW) for the five different \ion{Si}{2}~transitions (where available within the wavelength range of each individual spectrum) as a function of oscillator strength (commonly referred to as the ``curve of growth").  This allows us to search for the expected evolution from the optically thin to optically thick regime.  We find for many galaxies where all five transitions are visible in the spectrum, there is a clear transition from the optically thin portion of the curve of growth where the equivalent width evolves linearly with $\lambda f$, to the optically thick regime where the curve of growth flattens.  This is shown for our high signal-to-noise LBA spectrum in figure \ref{fig:curve_of_growth}.  Note that there is a transition of \ion{S}{3} at 1190.206$\rm \AA$ that contaminates the \ion{Si}{2} 1190.416$\rm \AA$ absorption line and probably explains why it unexpectedly appears stronger than \ion{Si}{2}~1193 in our figure though the latter has a stronger oscillator strength. This gives us confidence that for stronger absorption features of \ion{Si}{2}, (in particular, $\lambda=1260$) we can assume that the line is optically thick.  Thus, the relative residual intensity in the core of the absorption line can be interpreted as a measure of the fraction of the UV source not covered by optically thick low-ionization gas.  It should be noted of course that while in our model these sightlines are optically-thin to low-ionization metal lines, they could still be optically thick to ionizing radiation. Thus, observations below the Lyman edge are required to confirm the actual fraction of escaping ionizing radiation.

\subsection{Resonance and Fine Structure Line Emission}
\label{ssec:resonance}
As emphasized by \citet{Prochaska2011} and \citet{Scarlata2015}, the UV absorption lines we are using are resonance lines and as such, each upward transition resulting in absorption should be followed by a downward radiative transition resulting in emission. This emission will `infill' the absorption line, and can in principle have a significant effect on the net observed profile. Thus, correctly measuring properties such as optical depth, covering factor, and outflow velocity requires correcting the profiles for this infilling. 

Both of the above papers show that the fine-structure transitions that are associated with many of the commonly-used resonance lines can be used to deduce the properties of the infilling resonance line emission.  These fine structure lines share the same upper level as the associated resonance lines, but the lower level is slightly higher in energy than the ground state associated with the resonance line. Thus the downward radiative transition can occur via either the resonance line (resulting in infilling) or via the fine-structure line (which does not contaminate the observed absorption line). Quantitatively, the flux ratio between the resonance and fine-structure emission should be the ratio of the respective Einstein A coefficients.  

\begin{figure}
\includegraphics[scale=0.45,trim=50 20 0 510,clip=true]{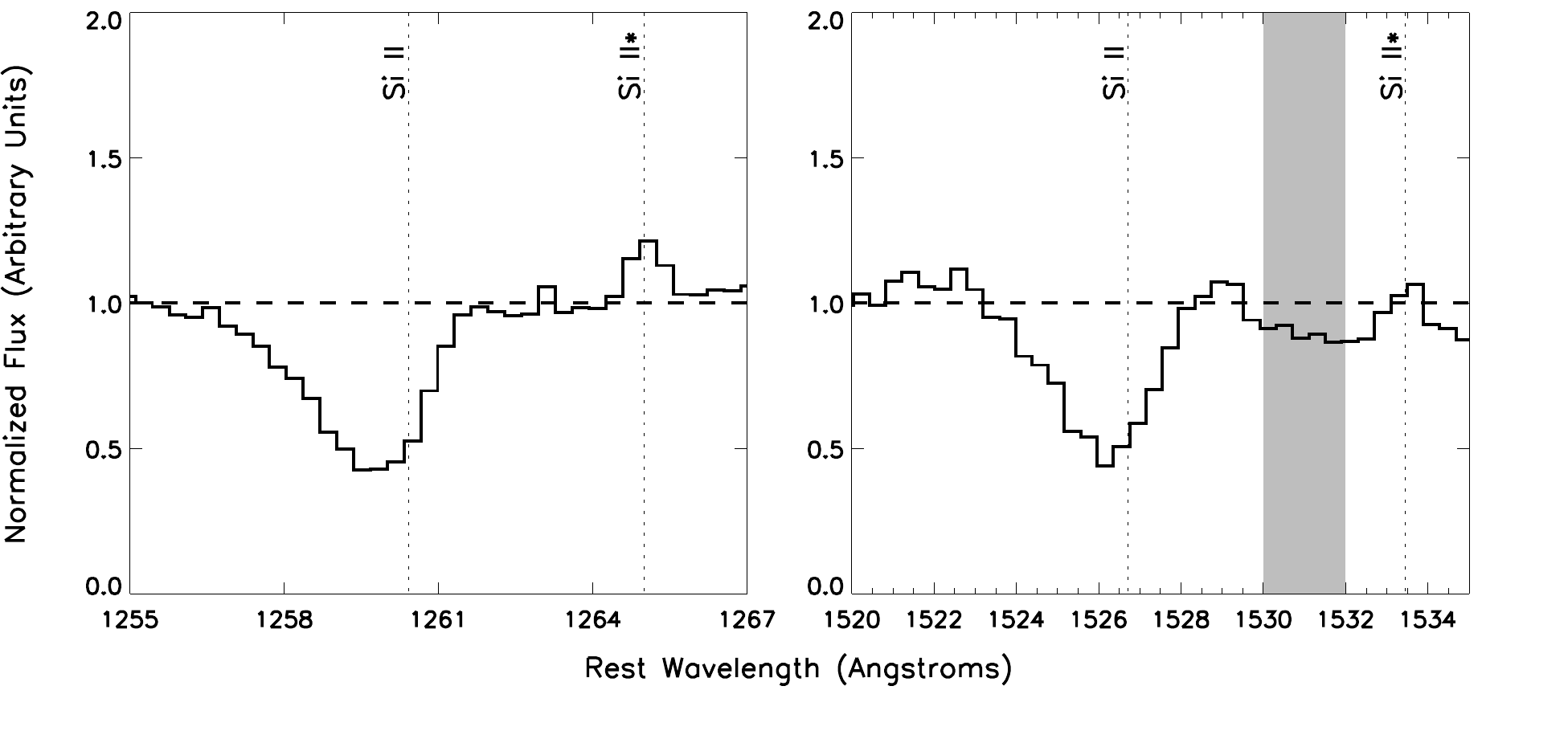}
\caption{\small Zoom-in on the spectral regions around Si II 1260.4 \& Si II* 1264.7 (left) and Si II 1526.7 \& Si II*1533.5 (right) from a combined spectra of our 22 LBAs.  Both absorption lines have associated resonance and fine-structure lines that can be used to estimate the amount of in-filling expected in the absorption line.  We measure respective ratios of the equivalent widths (resonance/fine-structure) of 0.10 and 0.11, implying that the respective ratios of the inferred (infilling) emission to observed absorption is only 0.09 and 0.006 and thus has a negligible effect on the properties of the gas inferred from the observed absorption lines.  We show in grey on the \textbf{right} where a local continuum was used to measure the equivalent width of Si II*1533.5 otherwise the inferred value of infilling emission would have been underestimated.}
\label{fig:scarlata}
\end{figure}

In Figure \ref{fig:scarlata} we show two pairs of resonance and fine-structure lines associated with Si II: Si II 1260.4 plus Si II* 1264.7 and Si II 1526.7 plus Si II*1533.5 in the coadded spectra of our 22 LBAs.  The ratios of the Einstein AÕs (resonance/fine-structure) for these two pairs are 0.87 and 0.051 respectively.   We measure respective ratios of the equivalent widths (resonance/fine-structure) of 0.10 and 0.11, implying that the respective ratios of the inferred (infilling) emission to observed absorption is only 0.09 and 0.006.  The fine-structure emission lines are narrow and centered on $v_{sys}$. It is therefore clear that infilling of the absorption lines will have a negligible effect on the properties of the gas inferred from the observed absorption lines.  

\subsection{\lya~Emission}
\label{ssec:lya}
\begin{figure*}
\includegraphics[scale=0.8]{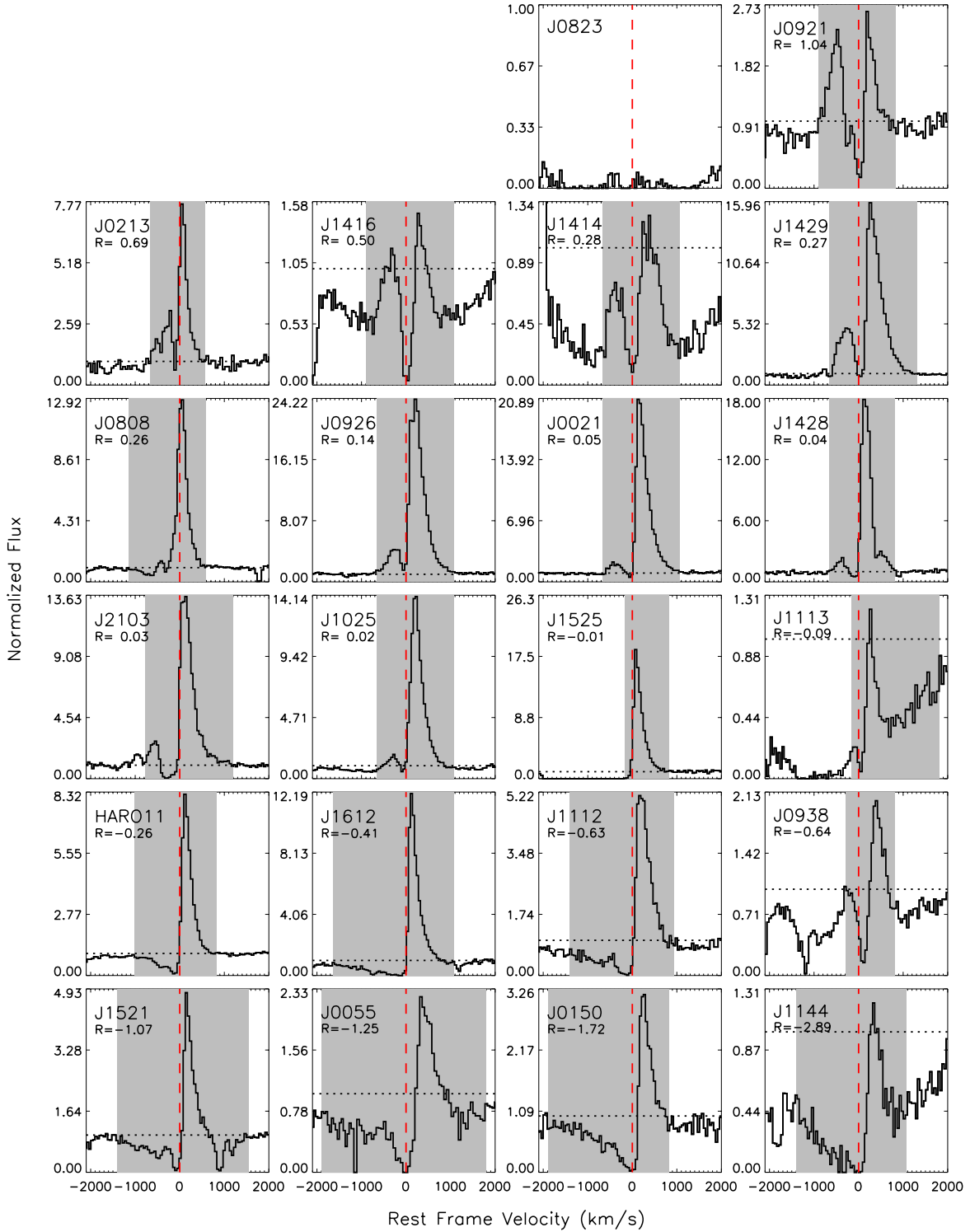}
\caption{\small Detailed \lya~profile of all LBAs in this sample.  Zero velocity is marked with a dashed red line.  The area over which values of $R_{eqw}$ and EW were determined are marked in grey.  These were chosen by eye.  For well-normalized data the continuum is expected to have a value around 1.0 (grey dotted line).  In the cases of J0055, J1113, J1144, J1414, and J1416 the global continuum fitting produced a poor fit around \lya~so a local continuum was estimated in a region close too \lya~and $R_{eqw}$ was calculated relative to this local value.  There is significant variety in the observed profile shapes.  While many objects display the traditional P-Cygni profile expected (resulting in a negative or small positive value of $R_{eqw}$), several show both redshifted and blueshifted emission resulting in a large positive value of $R_{eqw}$.}
\label{fig:lya}
\end{figure*}

H11 suggested that \lya~emission line profiles in which a significant fraction of the emission was blueshifted relative to the galaxy systemic velocity were indicative of galaxies in which there were holes in the distribution of outflowing neutral hydrogen. Indeed in our galaxies we notice a variety of different \lya~profile shapes.  Many galaxies display the traditional P-Cygni profile with redshifted emission and blueshifted absorption that are commonly seen in LBGs \citep{Shapley2003,Steidel2010}. However, other LBAs display profiles with an abundance of blueshifted \lya~emission. Following H11, we have quantified the different \lya~profiles shapes in the following way: we first measure the \lya~equivalent width blue-ward and red-ward of the systematic velocity, where we define the equivalent width for emission (absorption) to be positive (negative). We then take the ratio of the blue/red equivalent widths to parameterize the profile shape ($R_{eqw}$).  Therefore, a negative value of $R_{eqw}$ indicates blue shifted absorption and redshifted emission (a traditional P-Cygni profile) while large, positive values of $R_{eqw}$ indicate strong redshifted {\it and} blue shifted emission (see the comparison in figure \ref{fig:lya}).  It is worth noting that all but one of our objects shows strong \lya~emission, as opposed to some high redshift samples such as that of \citet{Shapley2003}.
 
 The \lya~equivalent width itself is of interest since it is a measure of the fraction of \lya~photons that escape the galaxy relative to the escaping fraction of FUV continuum photons. It is therefore plausible that the equivalent width of the \lya~emission line could be used as an indirect indicator of the escape of Lyman continuum photons. We have therefore measured a net \lya~equivalent width in emission for comparison with high-redshift samples.
 
For several objects (J0055, J1113, J1144, J1414, and J1416) the continuum normalization around \lya~was poor due to intervening absorption from either the circumgalactic medium of nearby galaxies or the local interstellar medium.  J1113 clearly shows evidence for a DLA at $z=0.1712$ associated with the CGM of J111324+293051 with metal lines that are all double-peaked and a strongly blue-shifted component that is due to the foreground galaxy also seen in the SDSS image at the same redshift.  In the other four cases, that of J0055, J1144, J1414, and J1416, we see only very weak \lya~emission-lines.  One possibility is that this is a result of errors in the global continuum fitting because there is no true continuum around \lya.  Figure \ref{fig:coadd} demonstrates that the spectrum around \lya~$\pm 25 \rm \AA$ or so could be affected by the broad \ion{N}{5} stellar wind line redward and the blending of absorption due to \lya, \ion{Si}{3} and \ion{N}{1} on the blue side.  Alternatively, this profile could also be due to underlying \lya~absorption in the same galaxy as seen in, for example, \citet{Schaerer2008} and \citet{Dessauges2010}.  

In either case, we choose to calculate $R_{eqw}$ and the equivalent width with a local continuum estimated in a region close to the \lya~emission line and fluxes were calculated relative to this local value.  If the broad \lya~absorption is produced by a foreground galaxy (external to the outflow) or by the dense interstellar medium of the galaxy (internal to the outflow), then we believe that measurements of the outflow component of \lya~should be made relative to the local continuum allowing the measurement to highlight the portion of \lya~emission that is escaping from the galaxy. In Figures \ref{fig:residual},\ref{fig:ranking} and \ref{fig:galaxyproperties} we show the objects where a local continuum was used in purple to highlight that they are not driving the observed correlations.  Measurements of $R_{eqw}$ using a global continuum could not be made in a meaningful way for comparison as it is not clear over what range the intervening absorption makes a significant contribution to the spectrum.  We find that these five measurements do not significantly affect our conclusions.

\subsection{Starburst Size and Morphology}
\label{ssec:dco}

All morphological information for our new sample of LBAs is taken from acquisition images for COS which are performed using the NUV MAMA detector. The UV half-light radius was determined using SExtractor to estimate the half-light radius from the Kron radius.  The measured half-light radii are listed in table \ref{tab:lba}.  For the previous sample UV half-light radii are measured from existing HST data taken with the Solar Blind Channel (SBC) on the Advanced Camera for Surveys (ACS) of HST (see O09 section 2 for more details) using SExtractor.

Most LBAs consist of a number of compact star-forming clumps with a range of sizes and brightnesses.  However, a minority consist of only a single, highly compact ($\sim$100 pc) and massive ($\sim 1 \times 10^9 M_{\bigodot}$) central source.  H11 and O09 have defined these ``dominant central objects" (DCOs) as those LBAs with extremely compact star bursting central regions as determined by UV photometry.  Any object where the ratio of UV flux in the central 0.2" to the flux within the FWHM of the COS PSF is greater than $\sim 25 \%$ is considered a DCO (O09; see Figure \ref{fig:postage1} and \ref{fig:postage2}).  This calculation was done for our Cycle 20 data using the Aperture Photometry Tool \citep{Laher2012} on the HST acquisition images to calculate the flux within each radius.  The results of this calculation can be found in table \ref{tab:lba}.  The classifications for the previous sample were taken from O09 and confirmed both visually and using APT.  H11 found a connection between those LBAs that were also DCOs and those objects that were strong candidates for escaping Lyman continuum photons hypothesizing that strong starbursts were capable of creating optically thin channels in the neutral ISM that would allow ionizing photons to escape the galaxy.      

\section{Ancillary Data}
\label{sec:ancillary}

\subsection{Star Formation Rates}
\label{ssec:halpha}

It was originally hypothesized in O09 that a systematically smaller SFR as calculated from \halpha~compared to SFR indicators from the FUV and infrared (IR) might indicate that ionizing radiation was escaping the galaxy.  While the FUV and IR luminosities together should be a robust tracer of the bolometric luminosity and SFR, escaping ionizing radiation might depress the measured value of the extinction-corrected \halpha~luminosity.

FUV luminosities are measured at rest-frame 1500$\rm \rm \AA$ from our COS data and then converted to 1600$\rm \rm \AA$ (not available in our dataset) assuming $f_{\lambda} \propto \lambda^{\beta}$ where $\beta$ is the spectral slope ($\beta = 2.32$log$_{10}(m_{FUV}-m_{NUV})$-2.0) and the FUV and NUV magnitudes are taken from GALEX.  This is a reasonable approximation in the FUV where the slope is expected to be roughly constant.  The FUV luminosity is then simply $L_{FUV}=\lambda L_{\lambda}$ at 1600$\rm \rm \AA$ \citep{Overzier2010}.  

For the H11 sample the IR luminosities are taken from \citet{Overzier2010} which uses SED fitting of infrared data obtained with the Infrared Array Camera (IRAC), the Multi-band Imaging Photometer (MIPS) and the InfraRed Spectrograph (IRS) on the Spitzer Space Telescope to estimate the IR luminosity, $L_{IR} (3-1000\mu m)$.  Unfortunately, none of the new sample in the current paper have been observed with Spitzer or Herschel, so our only infrared data comes from the Wide-Field Infrared Survey Explorer (WISE) All-Sky Survey where the W4 band is at an observed wavelength of 22$\mu m$.  This data is dominated by hot dust in the mid-IR whereas the IR luminosity is typically dominated by colder dust seen at longer wavelengths.  We have therefore determined an empirical relation between IR luminosity measurements based on Spitzer plus Herschel and measured W4 luminosities for the 28
 available LBAs (Overzier et al. in preparation). We use this relation to estimate the total IR luminosities in our new sample.  Finally, the $SFR_{FUV,corr}$ is calculated according to the prescription of \citet{Kennicutt2012} by first correcting the FUV luminosity with the IR luminosity and then converting to a SFR.

\begin{multline}
\log~SFR_{FUV,corr} ~(M_{\bigodot}~yr^{-1}) = \\ \log \left(L_{FUV,obs} + 0.46\times L_{IR,3-1000\mu m}\right (\nu L_{\nu}) ) - 43.35
\end{multline}

To measure the \halpha~luminosity the emission must first be corrected for dust extinction.  This is accomplished using the reddening law of \citet{Calzetti2000} for wavelengths greater than 6300$\rm \rm \AA$ to generate the correction factor,`f' :

\begin{equation}
f =10.0^{(0.4 \times E(B-V) \times 2.659(-1.857+10400/\lambda)+4.05)}
\end{equation}

where the extinction, $E(B-V)$ is calculated from the ratio of \halpha~to H$\beta$ flux as $E(B-V)_{gas} = {{\log_{10}[(f_{H\alpha}/f_{H\beta})/2.87]}\over{0.4 \times 1.17}}$.

To convert our extinction-corrected \halpha~luminosity to a SFR we use the same prescription as O09:
\begin{equation}
SFR_{H\alpha,0}[M_{\bigodot}~yr^{-1}] = 5.3 \times 10^{-42}L_{H\alpha,obs}[erg~s^{-1}]
\end{equation}

As in O09, an additional factor of $\sim 1.7$ is applied to account for the \halpha~flux outside the SDSS fiber.  We use the final indicator, $\log\left({{SFR_{FUV,corr}}\over{SFR_{H\alpha}}}\right)$ as possible evidence of escaping ionizing radiation.

\subsection{Measuring Stellar Masses}
\label{ssec:mass}
We use both a stellar mass for the entire galaxy and for the starburst. The total stellar mass is taken from the MPA-JHU galaxy catalog of SDSS DR7\footnote{Available from: http://www.mpa-garching.mpg.de/SDSS/DR7/}.  We fit a burst mass by normalizing the starburst mass of an instantaneous burst (or SFR yr$^{-1} \times$ age for a continuous burst) of our best-fit SB99 model for each galaxy to the flux from the UV spectrum. 

\subsection{Measuring Emission Line Diagnostics}
\label{ssec:bpt}
It was previously noted in O09 that some LBAs, were characterized by unusually weak [\ion{S}{2}]$\lambda$6717,6731 line emission compared to typical star-forming galaxies. Since in HII regions this line arises in gas at the outer boundary of the Stromgren sphere where the gas transitions from ionized to neutral (e.g. Pellegrini et al. 2012), relatively weak [SII] emission might indicate the presence of gas that is optically thin to ionizing radiation (e.g. density-bounded conditions).

\begin{figure*}
\includegraphics[scale=0.7,angle=-90,trim=50 20 0 0,clip=true]{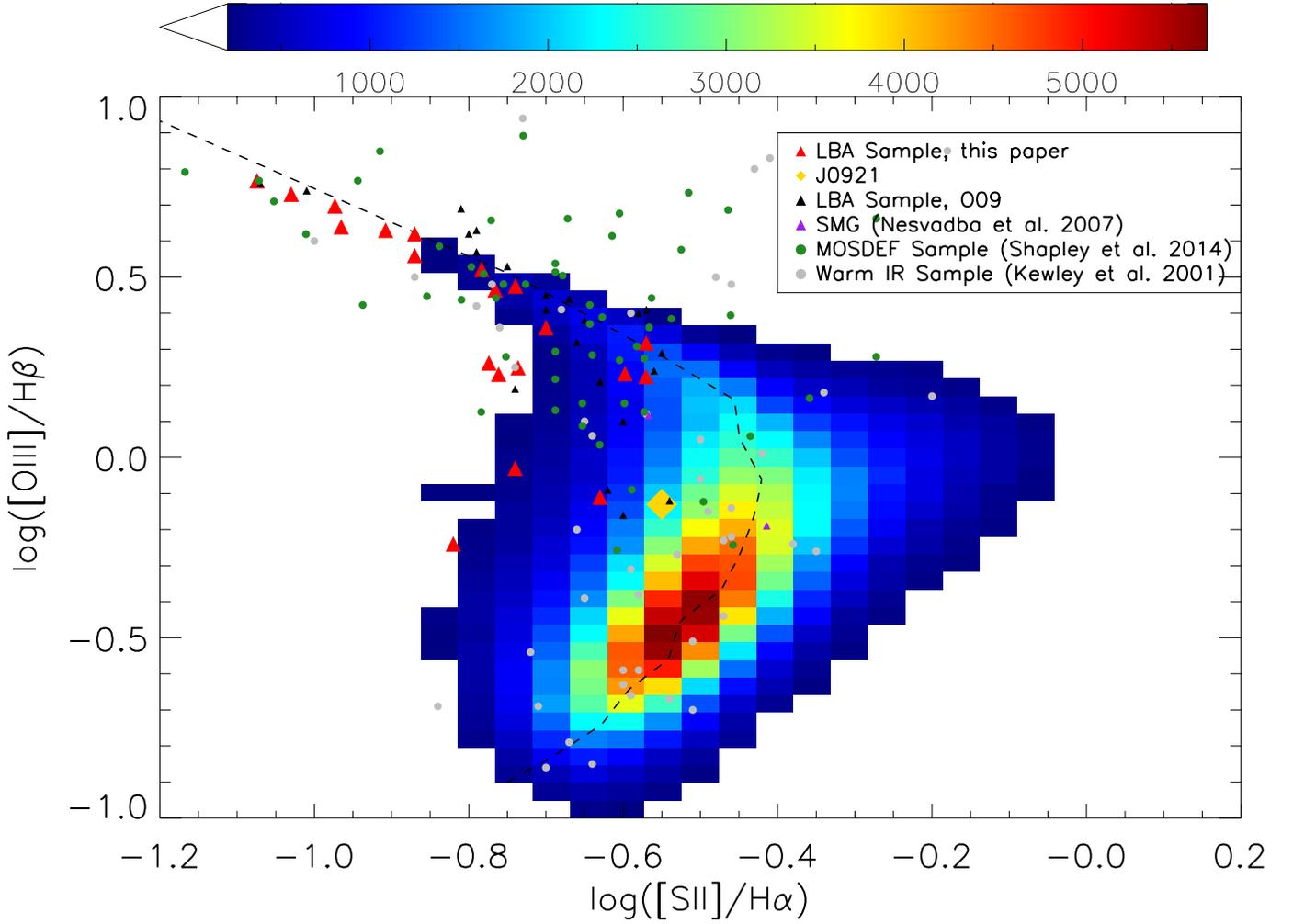}
\caption{\small Emission line ratio of log ([\ion{S}{2}]/\halpha) vs. $\log(\oiii/H\beta)$.  The density contours are SDSS galaxies categorized as star forming galaxies according to the emission line diagnostics of \citet{Kewley2006} in the  BPT diagram.  The dashed line is our empirically determined ``star-forming" ridgeline (see \ref{ssec:bpt}).  We calculated the perpendicular distance between our objects, shown in red, and the mode of star-forming galaxies on the diagram to express the deviation from a typical star-forming galaxy for this emission line diagnostic.  J0921, our object with confirmed Lyman continuum emission, is represented as a yellow diamond.  We also include some other samples for comparison.  The other LBAs observed with HST in O09 are shown in black.  We include the high-redshift SMG of \citet{Nesvadba2007} in purple as well as the high-redshift star-forming galaxies of the MOSDEF survey \citep{Shapley2014} in green.  Finally, we include a sample of nearby warm IR-luminous galaxies \citep{Kewley2001} in gray.}
\label{fig:bpt}
\end{figure*}

To quantify the relative strength of [\ion{S}{2}] emission in the LBAs, we first defined the locus of normal star-forming galaxies in the diagnostic diagram of log ([\ion{S}{2}]/\halpha) vs. $\log(\oiii/H\beta)$.  We took only those galaxies from SDSS that fell in the star-forming region of the $\log$ ([\ion{N}{2}]/\halpha) BPT diagram as defined by \citet{Kewley2006}.  We then followed a similar prescription to \citet{Kewley2006} to define the ridge-line.  We began with an empirical end-point at the bottom of the distribution of SDSS ``star-forming" galaxies and divided the sample in to radial bins of 0.1 dex.  For each radial bin we calculated the mode of the angle between the end-point and each galaxy in the bin.  This provided a parameterization of the ridge line as shown in Figure \ref{fig:bpt}.  The displacement value, $D_p$ is then defined to be the perpendicular distance between each LBA and the parametric ``ridge line".  It is evident in this figure that a number of LBAs tend to be significantly displaced from the ridge of normal SDSS star-forming galaxies.         

\section{Results}
\label{sec:results}

\subsection{Comparing Indirect Indicators of Lyman Continuum Escape}
\label{ssec:rank}
Following O09 and H11, we consider three parameters as being indirect indicators of escaping Lyman continuum photons:

\begin{enumerate}
  \item The amount of residual intensity in the core of the \ion{Si}{2}~$\lambda 1260$ ISM absorption line (\%)
  \item The shape of the \lya~line emission line profile ($R_{eqw}$). 
    \item The distance from the star-forming ridge line in the space of log ([\ion{S}{2}]/\halpha) vs. $\log(\oiii/H\beta)$ (dex)
\end{enumerate}

\begin{figure*}
\includegraphics[scale=0.35,angle=90, trim=10 15 100 20,clip=true]{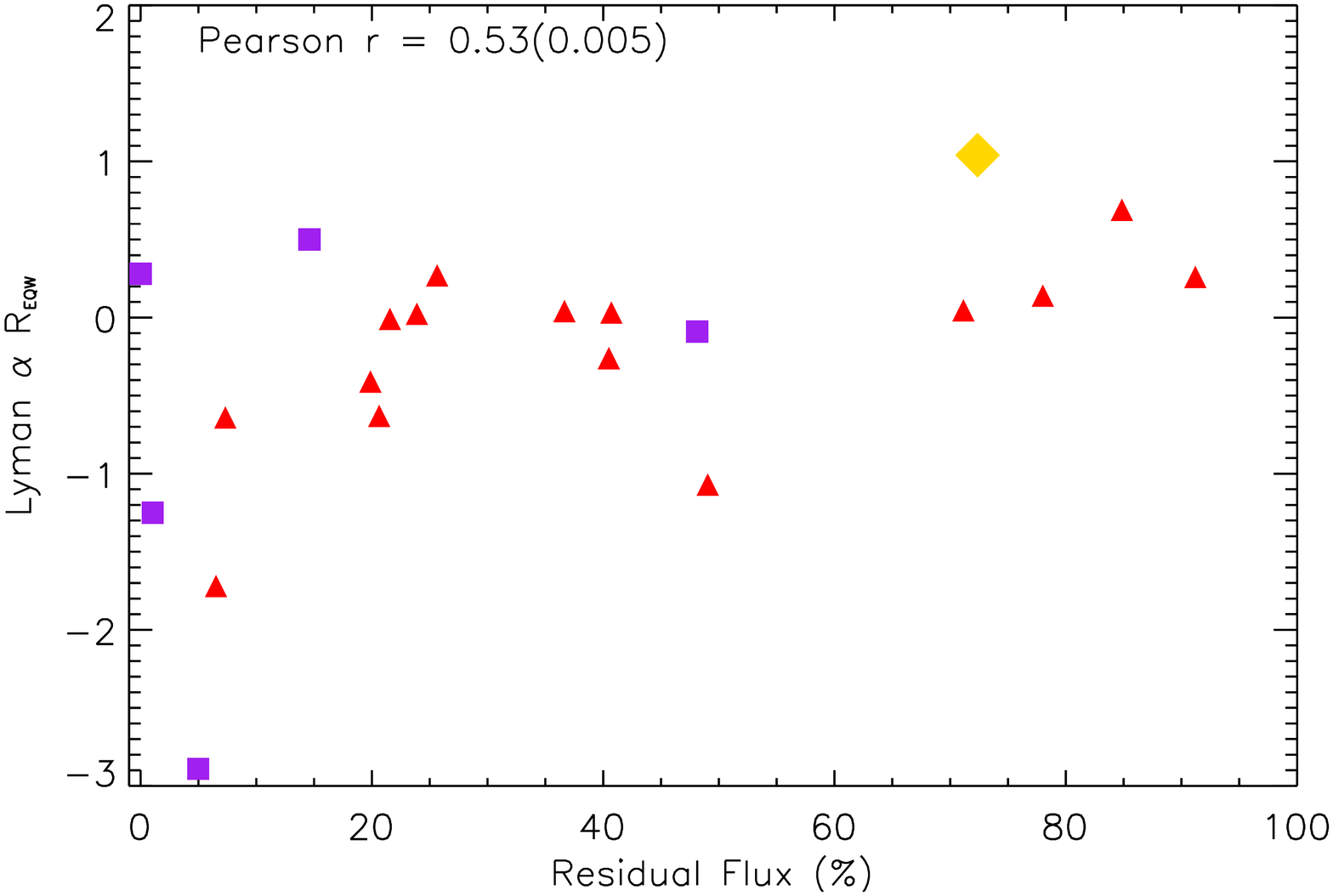}
\includegraphics[scale=0.35,angle=90, trim=10 15 100 20,clip=true]{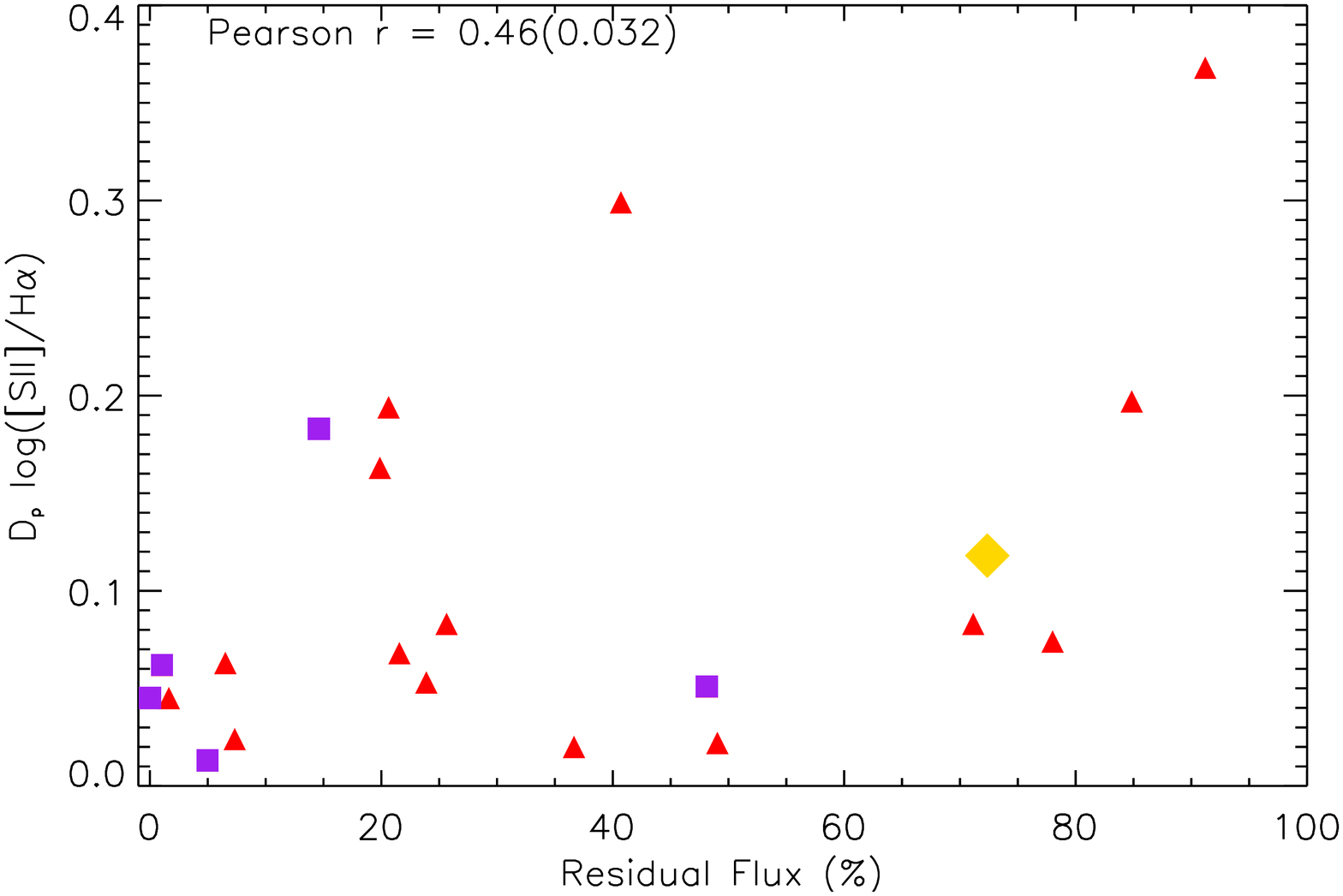}
\includegraphics[scale=0.35,angle=90, trim=10 15 100 20,clip=true]{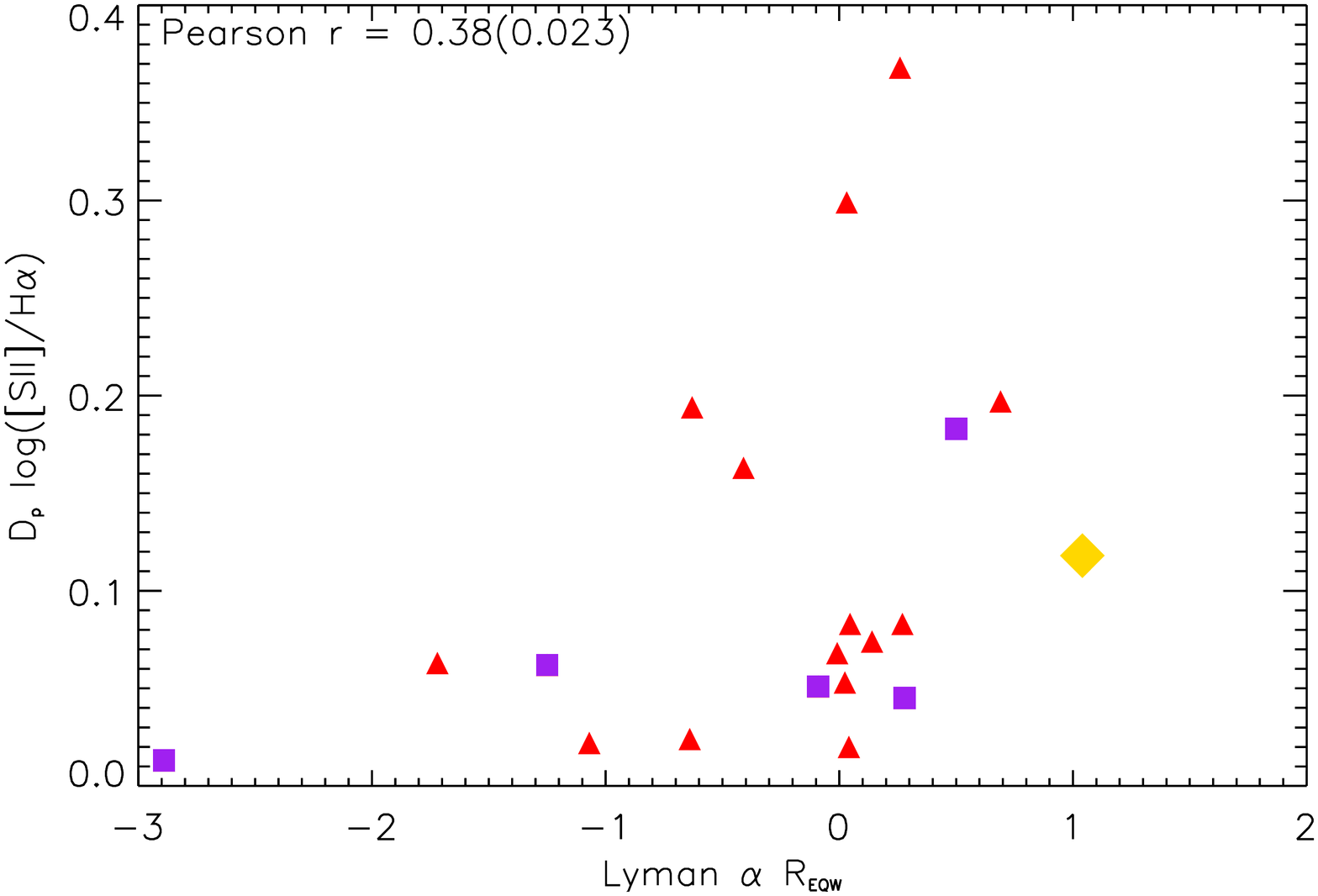}
\caption{\small Comparison of our indirect indicators of escaping Lyman continuum flux. Top, left: comparison of residual flux in the \ion{Si}{2}~absorption line (\%)  to the net equivalent width of \lya~(ratio of equivalent width blueward:redward of the \lya~rest wavelength). Top, right: comparison of residual flux in the \ion{Si}{2}~absorption line (\%) to the distance from the star-forming ridge line in log (\ion{S}{2}/\halpha) vs. $\log(\oiii/H\beta)$ space (dex).  Bottom, left: comparison of the net equivalent width of \lya~(ratio of equivalent width blueward:redward of the \lya~rest wavelength) to the 
distance from the star-forming ridge line in log (\ion{S}{2}/\halpha) vs. $\log(\oiii/H\beta)$ space (dex).  J0921, our object with confirmed Lyman continuum emission \citep{Borthakur2014}, is represented as a yellow diamond.  Objects for which a local continuum fit to \lya~was used are represented here as purple squares. The Pearson r correlation coefficient is given in the top left corner of each figure with statistical significance of the deviation from zero in parentheses.}
\label{fig:residual}
\end{figure*}

The calculated values for each indicator measured in every LBA from our sample are presented in table \ref{tab:lba}.  When we compare each of these indicators, we find statistically significant correlations between all of them (see Figure \ref{fig:residual}).  We see a regime with high residual flux (indicating incomplete coverage of the starburst by optically-thick low-ionization gas), the presence of a significant amount of blue-shifted \lya~photons escaping the galaxy (indicating incomplete coverage of the near-side of the outflow by neutral hydrogen), and a relative deficit of [\ion{S}{2}] emission (indicating the presence of portions of the ionized gas that are optically-thin to the Lyman continuum). As previously noted by H11, these indicators tend to be most pronounced in those objects classified as DCOs with strong compact star forming cores. We will reassess this quantitatively in the next section.  We do not find a strong correlation between these indicators and the ratio of SFRs derived from the FUV luminosity vs. the \halpha~luminosity (see section \ref{ssec:other}) and so have excluded it here as an indicator of escaping Lyman continuum radiation.

\begin{figure*}
\includegraphics[scale=0.8,trim=40 30 0 500,clip=true]{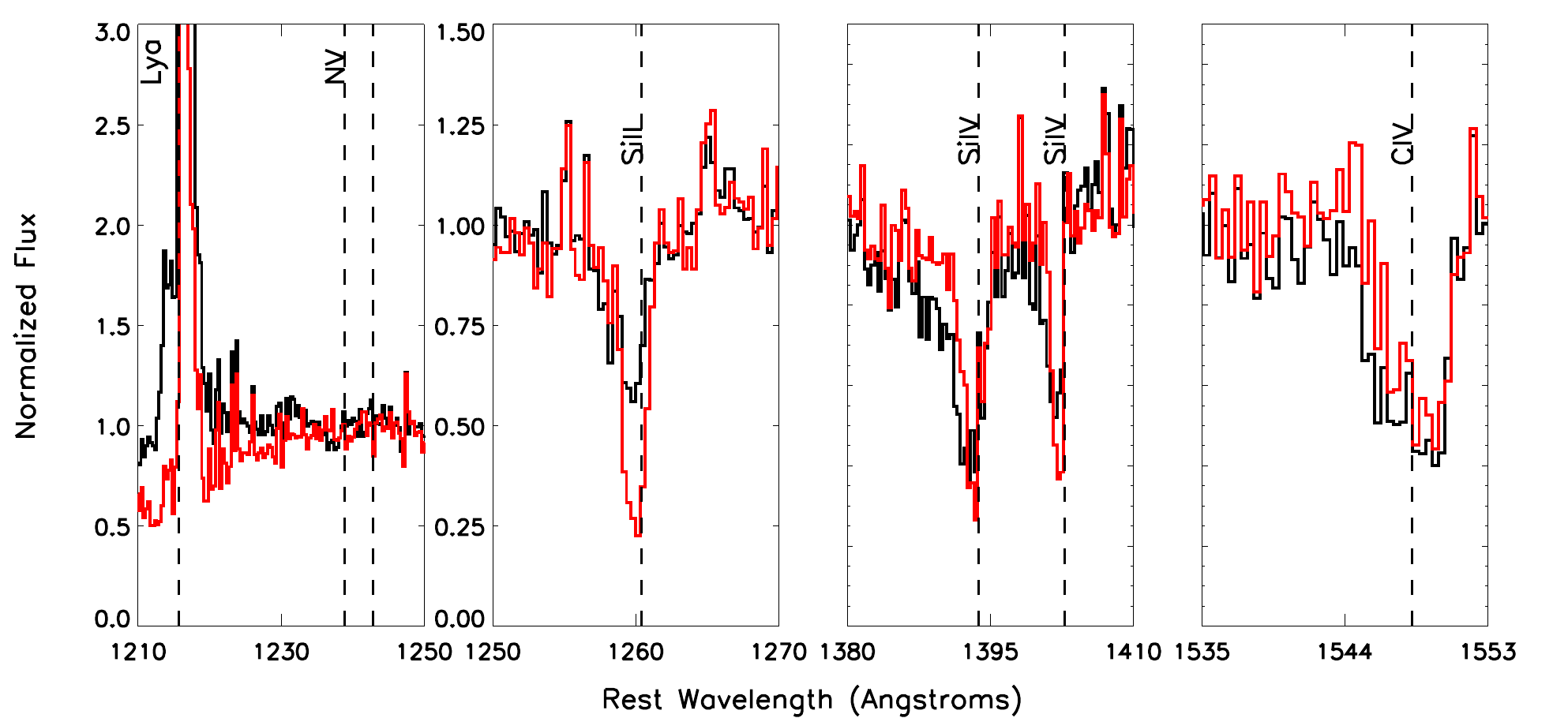}
\caption{\small Coadd of the 11 most ``leaky" objects (black) and the 11 least ``leaky" objects (red).  No weighting of individual spectra was applied. All spectra have been moved to the same wavelength grid as the SB99 models and the best-fit stellar model subtracted.  Important lines are noted.  }
\label{fig:coadd_compare}
\end{figure*}

Based on these results, we rank each of our galaxies in each of the three above parameters from the galaxy having the strongest indication of escaping ionizing radiation to the one with weakest indicator.  Thus, galaxies with a large residual flux, positive $R_{eqw}$, and a large distance from the star forming ridge in log ([\ion{S}{2}]/\halpha) vs. $\log(\oiii/H\beta)$ space will have the first rank in each category. Then we average the rank in all three categories to order our LBAs from most (1) to least (22) likely to show evidence for escaping ionizing radiation.  In one of our LBAs (J0921) - ranked ``third" in our scale of indirect Lyman continuum escape indicators - follow-up observations with HST COS have indeed revealed a dust-free escape fraction of $21\%$, the highest measured in any local galaxy \citep{Borthakur2014}.  Thus, our indirect indicators do (at least in this case) correctly predict the presence of real escaping ionizing flux.  The differences in key absorption features for our ``leaky" and ``non-leaky" candidates are best shown in figure \ref{fig:coadd_compare} where we have combined the spectra of the ``leakiest" half of the sample as well as the least ``leaky" half of the sample.  It is easy to see evidence of the parameters we selected on, e.g. the ``leaky'' sample shows much weaker Si II absorption at 1260 $\rm \AA$ but we also see evidence of other emergent properties e.g. the equivalent width of 
\lya~is smaller in the less ``leaky'' galaxies.

\subsection{Comparison with Galaxy Properties}
\label{ssec:other}

Given these results, we now compare our ``leaky indicator" rank to other galaxy properties to look for those that correlate with the likely escape of ionizing radiation.  For a comparison of the correlation between individual indirect indicators of escaping Lyman continuum radiation and galaxy properties see Table \ref{tab:pearson}. 

\begin{figure*}
\includegraphics[scale=0.35,angle=90, trim=10 15 100 20,clip=true]{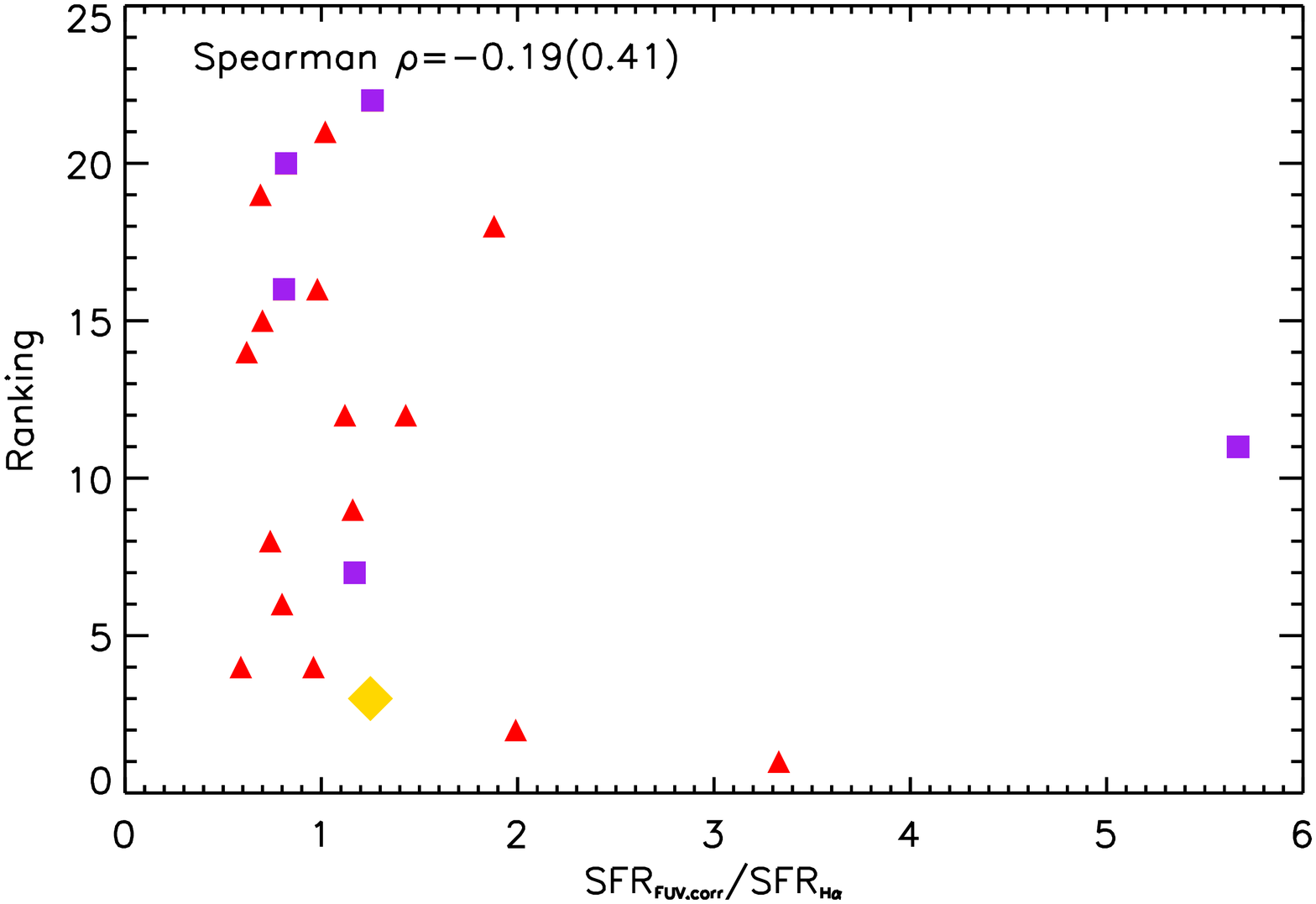}
\includegraphics[scale=0.35,angle=90, trim=10 15 100 20,clip=true]{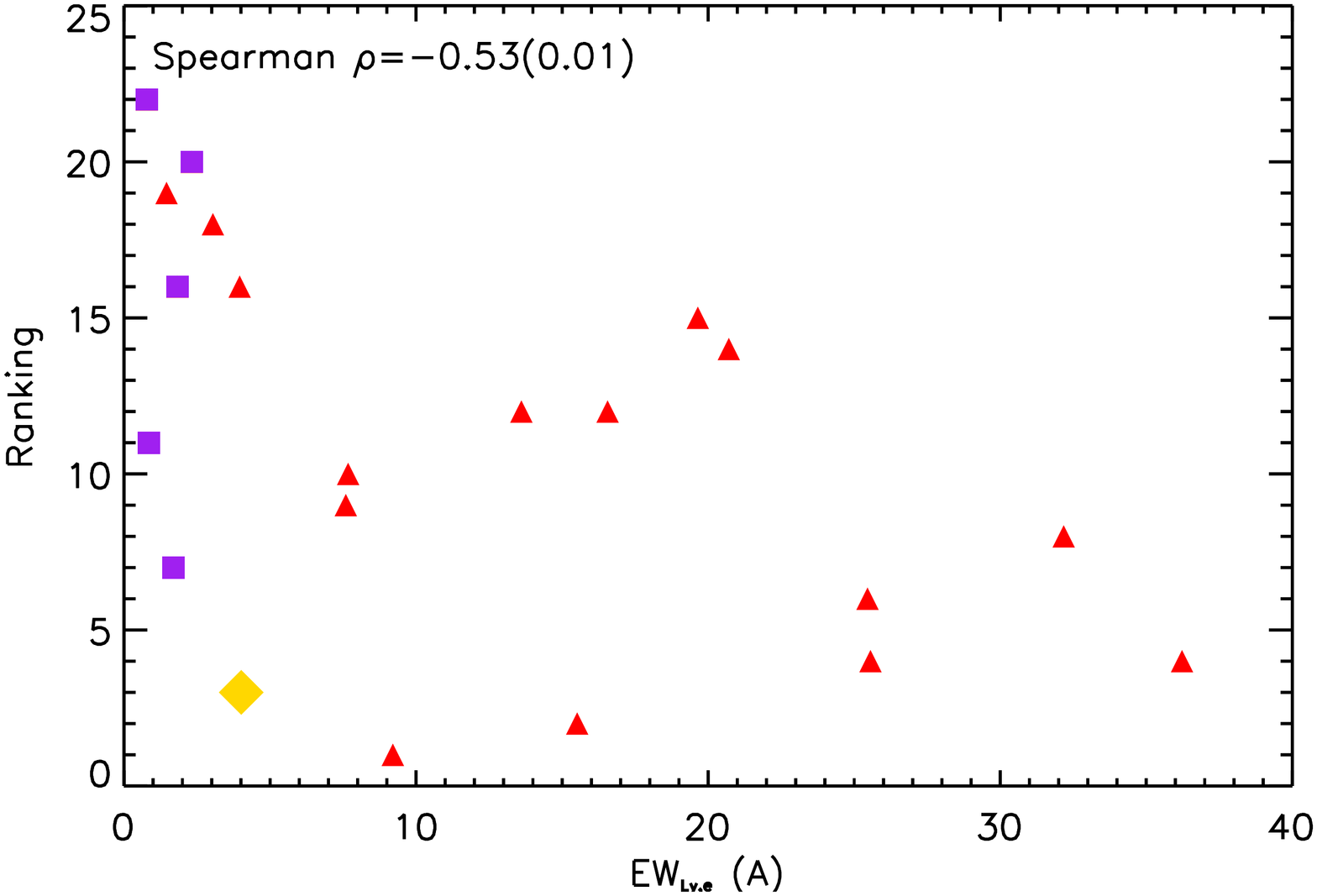}
\includegraphics[scale=0.35,angle=90, trim=10 15 100 20,clip=true]{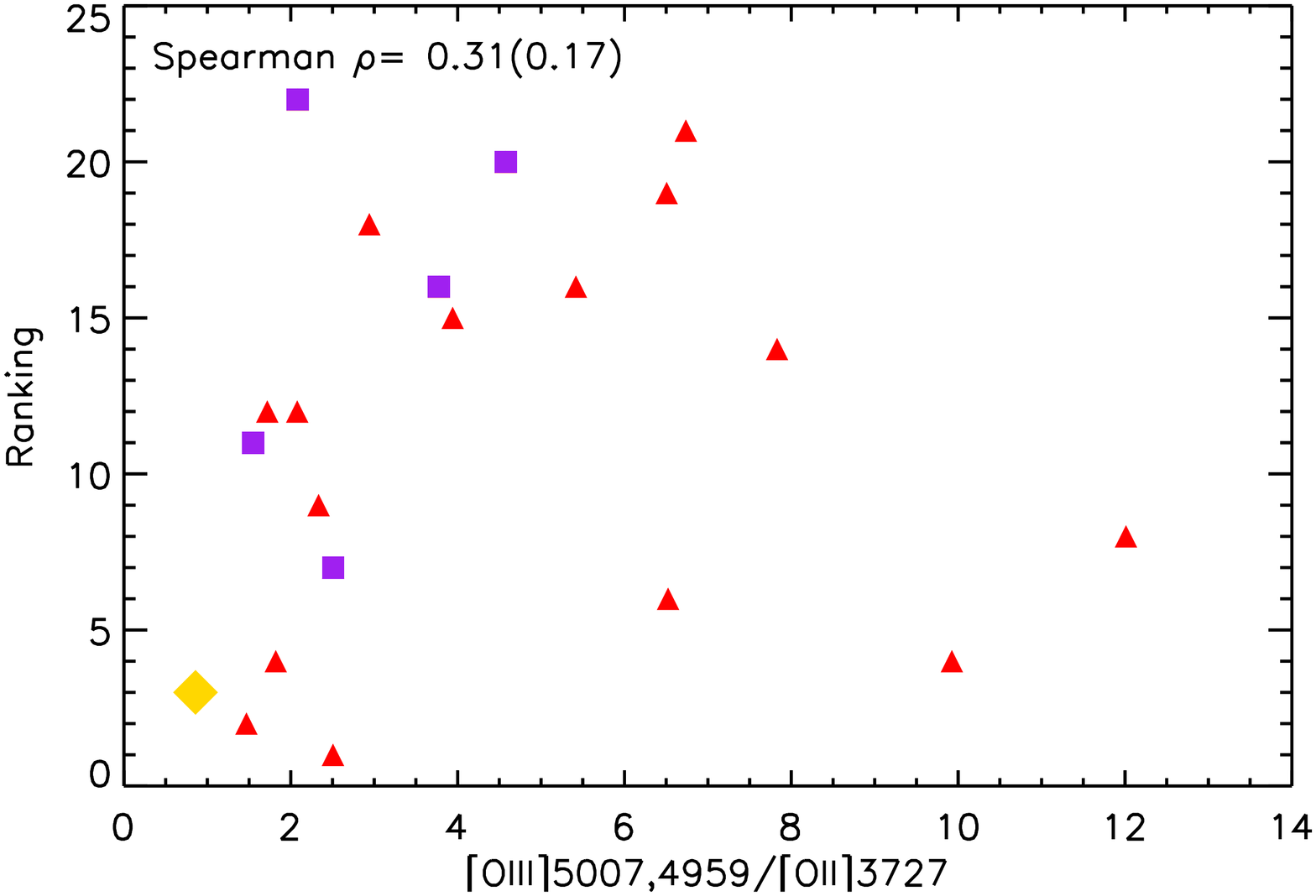}
\caption{\small Correlation between our leaky galaxy candidate ranking and various properties expected to correlate with leaky Lyman continuum.  The top left of each figure lists the calculated Spearman rank correlation coefficient for each relation as well as the statistical deviation from zero (in parentheses).  J0921, our object with confirmed Lyman continuum emission  \citep{Borthakur2014}, is represented as a yellow diamond.  Objects for which a local continuum fit to \lya~was used are represented here as purple squares.  1 (top, left): Ratio of SFR derived from IR-corrected FUV luminosity to extinction-corrected \halpha~luminosity. We find only a weak correlation between this value and our ``leakiness" ranking. 2 (top, right):  The equivalent width of \lya~(measured in emission only) compared to our leaky galaxy rank.  We find a strong correlation between the \lya~equivalent width and our ranking. 4 (bottom, left): The ratio of \oiii ~$\lambda5007,4959$ flux to \oii ~$\lambda3727$ emission to our leaky candidate ranking.  A deficit of \oii is often used to identify potential leaking galaxies but we find an anti correlation between this indicator and our leaky candidate ranking.}
\label{fig:ranking}
\end{figure*}

We begin by evaluating three other parameters that have been proposed to be indirect indicators of leakiness. O09 and H11 suggested that the ratio of the SFR derived from dust-corrected FUV luminosity to that derived from the dust-corrected \halpha~luminosity was such an indicator. Jaskot \& Oey (2013) proposed using the flux ratio of the [OIII]$\lambda$5007/[OII]$\lambda$3727 emission lines, and numerous authors \citep{Jaskot2014,Shibuya2014} have proposed using the equivalent width of the \lya~emission line as a diagnostic. We show the correlations of each of these with our ``leakiness rank'' in Figure \ref{fig:ranking}. We see no significant correlation of rank with either the SFR ratio or [OIII]/[OII].  We believe in this first case this is probably because the intrinsic scatter in these SFR indicators is higher than the escape fraction in these galaxies. On the other hand, there is a significant correlation between rank and \lya~equivalent width, albeit with exceptions like J0921 which has relatively weak \lya~emission.

\begin{figure*}
\includegraphics[scale=0.32,angle=90]{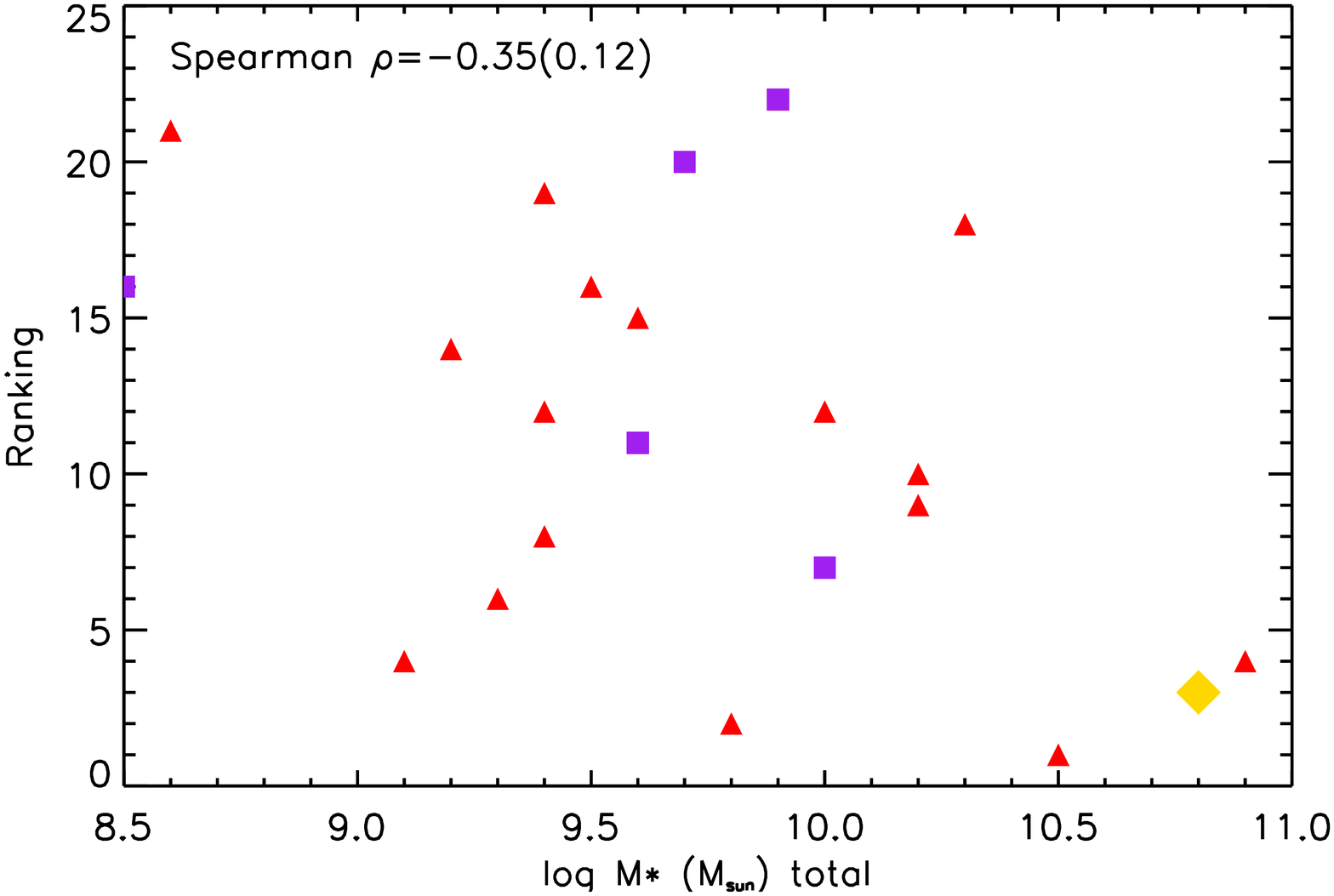}
\includegraphics[scale=0.32,angle=90]{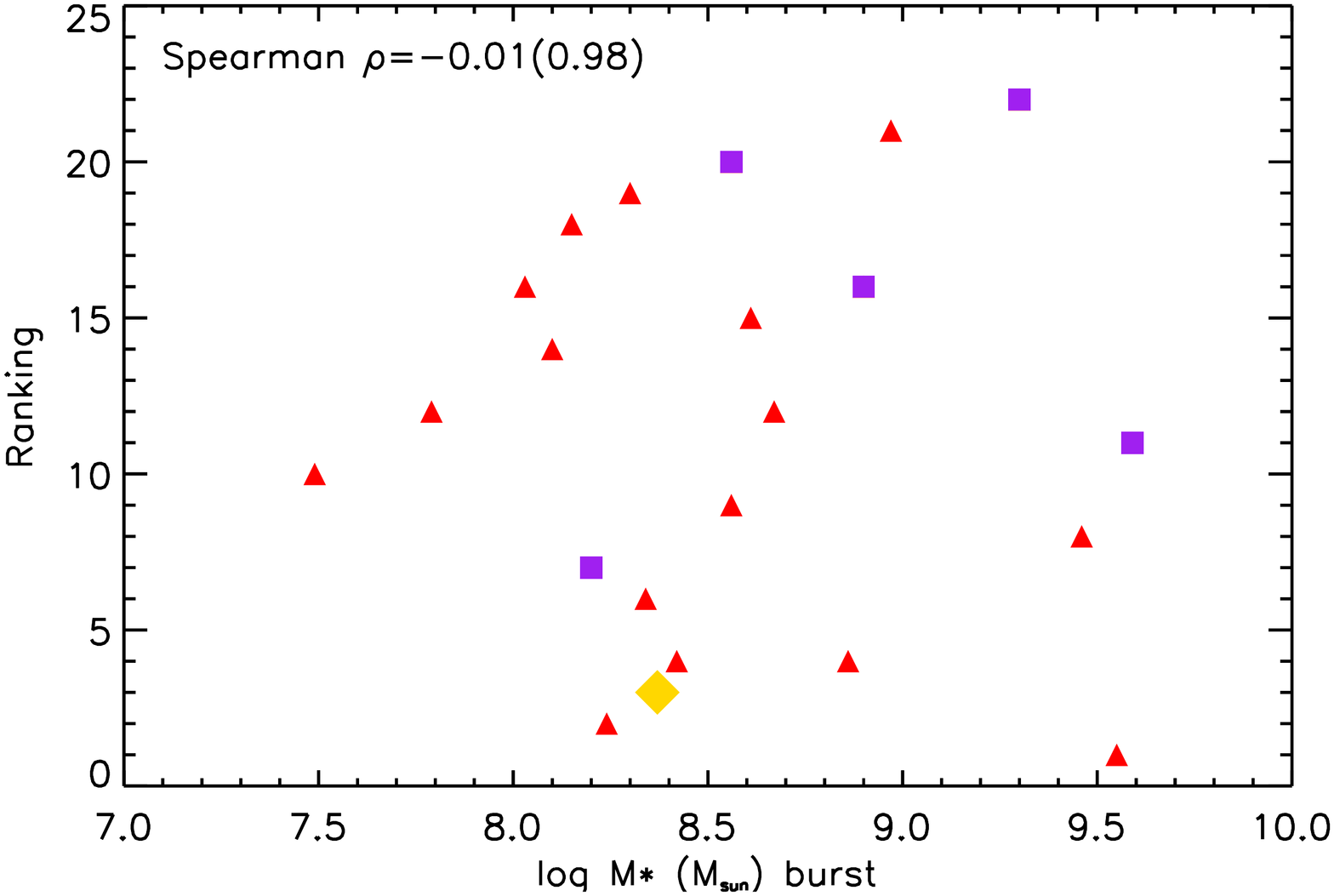}
\includegraphics[scale=0.32,angle=90]{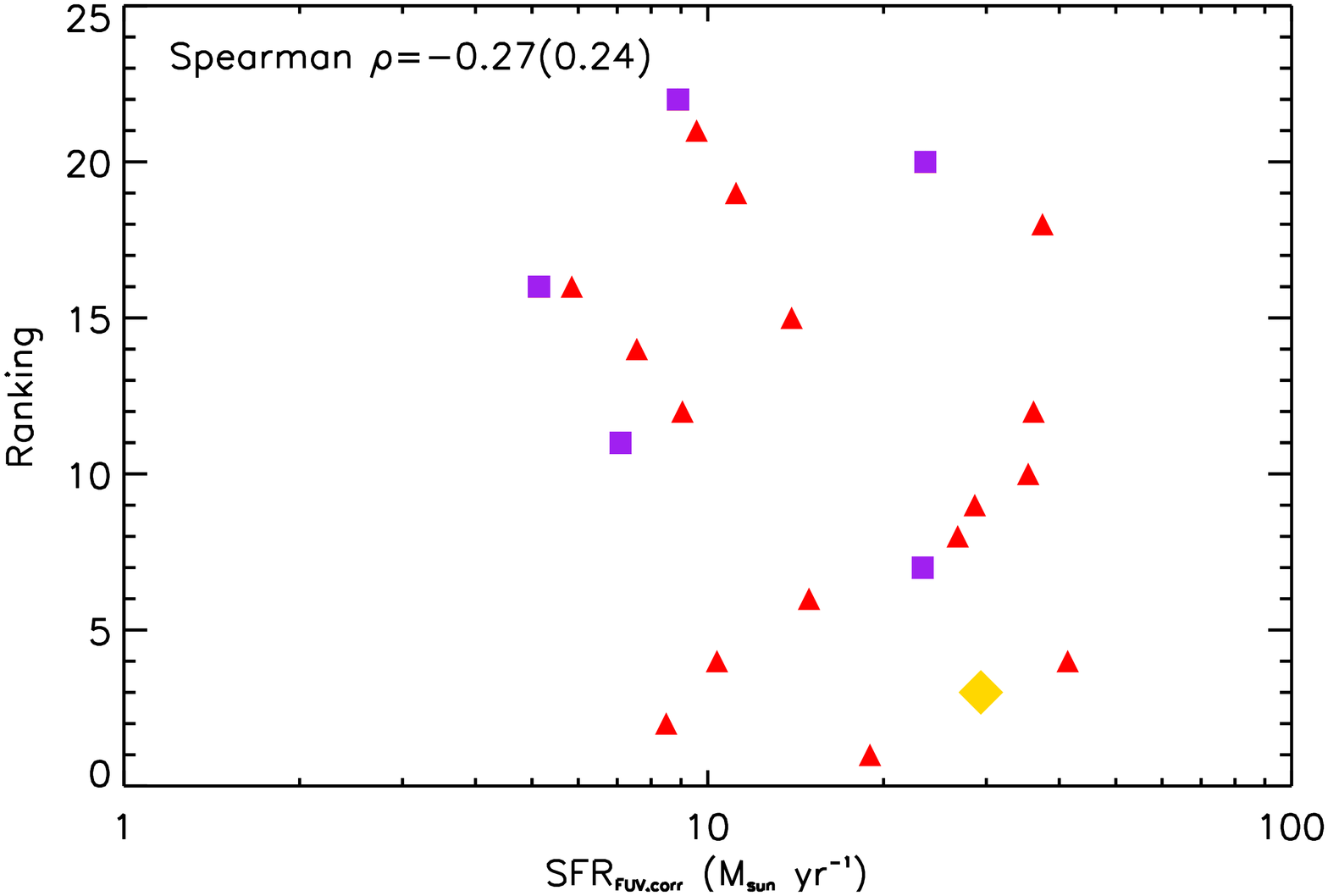}
\includegraphics[scale=0.32,angle=90]{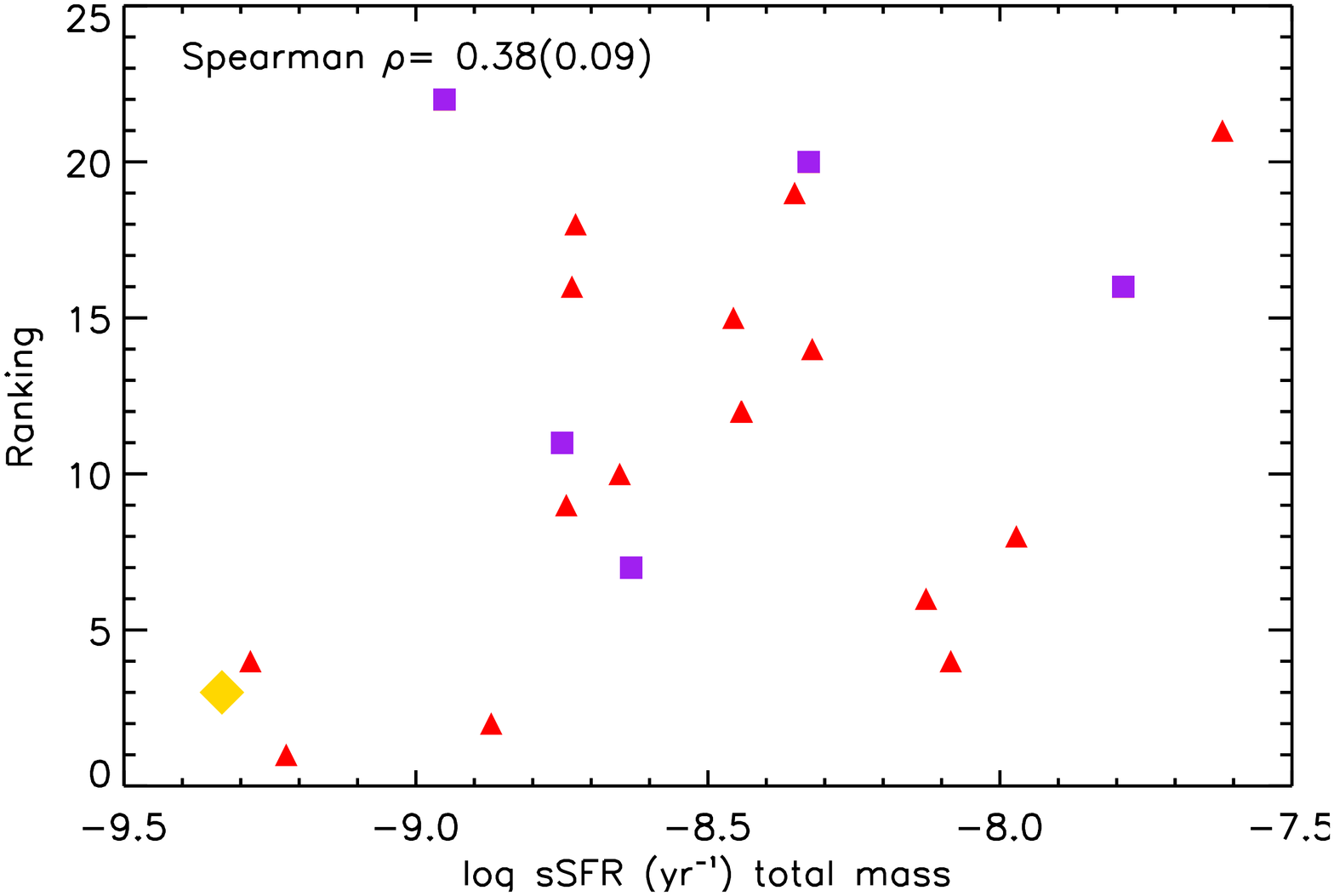}
\includegraphics[scale=0.32,angle=90]{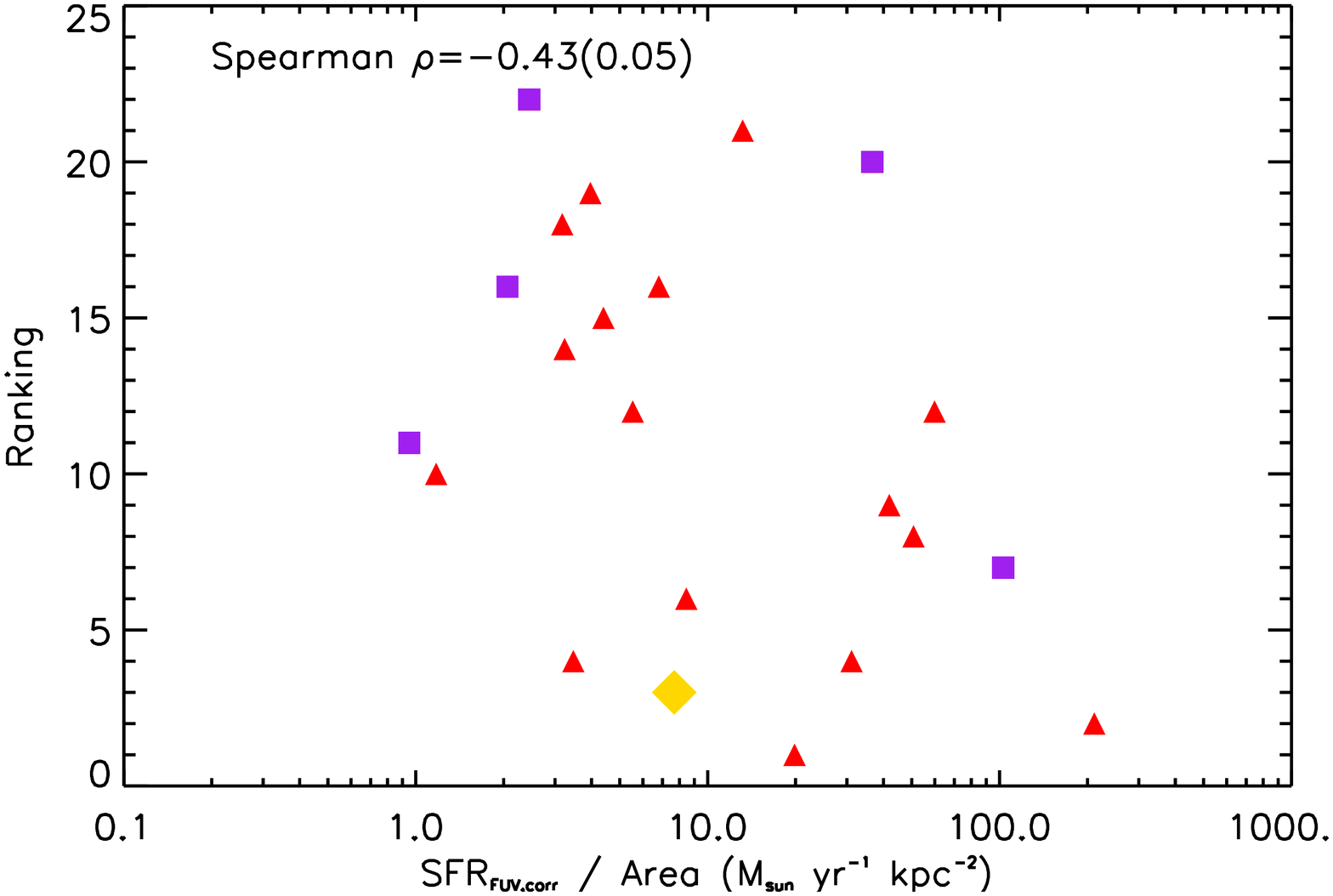}
\includegraphics[scale=0.32,angle=90]{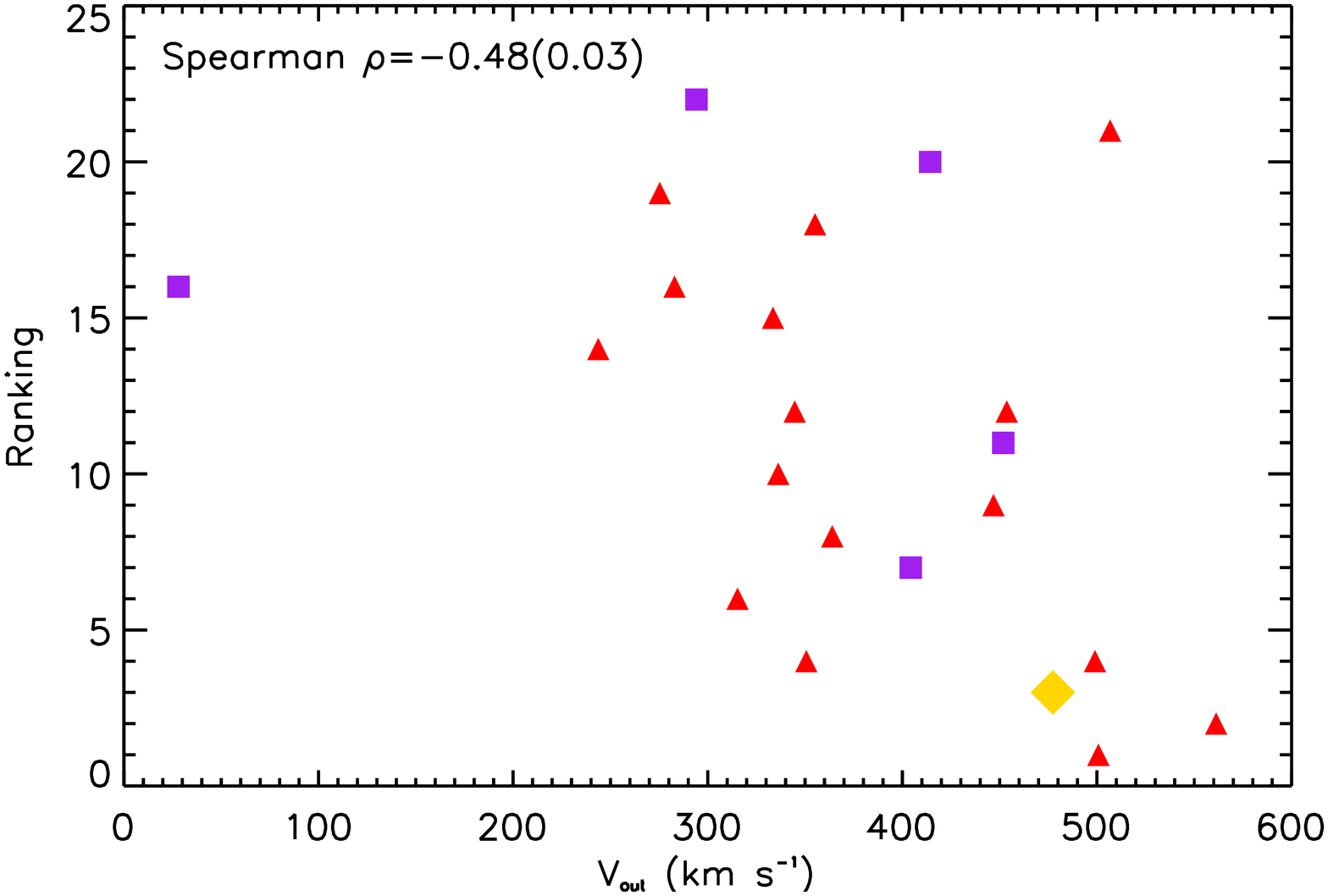}
\caption{\small Correlation between our leaky galaxy candidate ranking and various galaxy properties.  The top left of each figure lists the calculated Spearman rank correlation coefficient for each relation as well as the statistical deviation from zero (in parentheses).  J0921, our object with confirmed Lyman continuum emission  \citep{Borthakur2014}, is represented as a yellow diamond.  Objects for which a local continuum fit to \lya~was used are represented here as purple squares.  1 (top, left): Total galaxy mass as measured by the MPA-JHU catalog. 2 (top, right): Total burst mass as measured from our best-fit SB99 models.  3 (middle, left): Star formation rate derived from IR-corrected FUV luminosity for each of the LBAs.  4 (middle, right): Specific star formation rate (star formation per unit total stellar mass) for each LBA. 5 (bottom, left): Star formation rate per unit area for each of our LBAs.  6(bottom, right): Outflow velocity measured from the flux-weighted centroid of the \ion{Si}{3}~absorption line. The strongest trend shows that leakier objects are likely to reside in galaxies with compact star-forming regions and high outflow speeds.}
\label{fig:galaxyproperties}
\end{figure*}

Next, we examine how the leakiness rank correlates with the basic physical and dynamical properties of the galaxies.  In order of the strength of the correlation, we find that galaxies which are most likely to be ``leaky'' tend to have significantly higher star formation rates per unit area and higher measured outflow speeds (see Figure \ref{fig:galaxyproperties}). We do not find significant correlations between leakiness rank and either the SFR or the burst mass.

\begin{deluxetable}{|l|r|r|r|}
\tabletypesize{\scriptsize}
\tablewidth{0pt}
\tablecolumns{4}
\tablecaption{Pearson correlation between escape indicators and general LBA characteristics (significance of deviation from zero).\label{tab:pearson}}
\tablehead{\colhead{Indicator}&\colhead{EW\tablenotemark{1}}&\colhead{${{SFR_{FUV}} \over {area}}$\tablenotemark{2}}&\colhead{v$_{out}$\tablenotemark{3}}}
\startdata
F$_{res}$\tablenotemark{4}&0.42(0.026)&0.28(0.153)&0.41(0.031)\\
$R_{eqw}$\tablenotemark{5}&0.42(0.021)&0.22(0.142)&0.20(0.171)\\
$D_{\sii}$\tablenotemark{6}&0.18(0.236)&0.77(0.001)&0.63(0.001)\\
\enddata
\tablenotetext{1}{{Lyman} $\alpha$ equivalent width (EW) measured over the same region as $R_{eqw}$ but calculated using only emission.  More details can be found in section \ref{ssec:lya}}
\tablenotetext{2}{Star formation rate ($M_{\bigodot}$ yr$^{-1}$) calculated from the FUV luminosity corrected with the observed far-Infrared luminosity according to \citet[][For more information see section \ref{ssec:halpha}]{Kennicutt2012} divided by the area of the galaxy inside the petrosian radius enclosing 50$\%$ of object flux in the UV (kpc; for additional details see section \ref{ssec:dco})}
\tablenotetext{3}{Outflow velocity calculated as the first moment of the \ion{Si}{3}~absorption line (km s$^{-1}$; see section \ref{ssec:velocity} for additional information).}
\tablenotetext{4}{Residual Flux in the \ion{Si}{2}~absorption line measured at the centroid of the \ion{Si}{2}~1260 $\rm \AA$ line (\%).  For a description of the measurement see section \ref{ssec:res_flux}.}
\tablenotetext{5}{{Lyman} $\alpha$ equivalent width ratio of blueshifted/redshifted equivalent width where a negative value is absorption and a positive value is emission.  More details are provided in section \ref{ssec:lya}}
\tablenotetext{6}{Perpendicular distance from the star-forming ridge line measured on a plot of log (\ion{S}{2}/\halpha) vs. $\log(\oiii/H\beta)$(dex).  For additional details refer to section \ref{ssec:bpt}}
\end{deluxetable}

\section{Discussion}
\label{sec:discussion}

\subsection{Relation Between \lya~Emission and Escaping Ionizing Radiation}

There have been increased efforts to model the expected \lya~profile shape for different possible geometries and kinematic models for the interstellar medium. Much of the motivation has been to attempt to relate the properties of the \lya~emission line to the escape of ionizing radiation. Simpler models, such as those of \citet{Steidel2010} and \citet{Erb2010} already provide insight into the possible origins of our \lya~line profiles.  A classical P-Cygni profile is the product of a spherical outflow such as those produced in the winds of early-type stars \citep[e.g.][]{Castor1970,Castor1979} or starburst galaxies \citep[e.g.][]{Ahn2000,Verhamme2006}.  We see blueshifted absorption from outflowing gas approaching the observer on the front side of the galaxy.  Meanwhile, \lya~photons are able to resonantly scatter into our line of sight and escape the galaxy most effectively by scattering off outflowing material moving away from the observer on the back-side side of the outflow. These redshifted photons can then successfully traverse the HI on the front-side, which we see as redshifted emission.  A low column density of neutral hydrogen gas in the galaxy (and thus a small $\tau_{Ly\alpha}$ at the systematic velocity of the galaxy and in outflows) in a spherically symmetric outflow would allow \lya~photons to escape through an optically-thin medium even in outflowing gas on the side of the galaxy near the observer and produce blueshifted emission instead.  This idea was applied by \citet{Erb2010} to Q2343-BX418, a young, dust-free low-metallicity galaxy at $z = 2.3$, which displays significant blueshifted emission in the \lya~line.  

Much effort has been put into creating more complicated radiative transfer models of \lya~photons in an expanding medium.  Most recently, \citet{Behrens2014,Duval2014,Verhamme2014} find that, for an expanding shell with holes in the neutral gas, a \lya~profile with a main peak at the systematic redshift of the object as well as emission blueward of the systematic redshift are good indicators of escaping ionizing continuum radiation-- both characteristics seen in our \lya~profiles where we predict a high escape fraction of ionizing radiation. While symmetric \lya~profiles could also be due to radiative transfer in a static medium \citep[][e.g.]{Neufeld1991, Verhamme2006} this possibility is ruled out by the high outflow speeds we measure in the interstellar  absorption lines.  Most radiative transfer models show a secondary, redshifted peak in \lya~from emission in an expanding shell which we see blended with the central peak of photons escaping directly into our line of sight from the clumpy ISM.

Most importantly perhaps, we do find a statistically significant correlation between our three indirect indicators of escaping ionizing radiation and the equivalent width of the \lya~emission line.  A high equivalent width is often used as a line of evidence for escaping ionizing continuum radiation \citep{Jaskot2014,Shibuya2014}.  Given the prominence of the \lya~emission line and the many on-going and planned surveys of \lya~emitters at high-redshift, this is result is a promising one for understanding re-ionization.     

\subsection{The Relation Between Starburst Compactness and Escaping Ionizing Radiation}

\begin{figure}
\includegraphics[scale=0.32,angle=90, trim=10 15 100 20,clip=true]{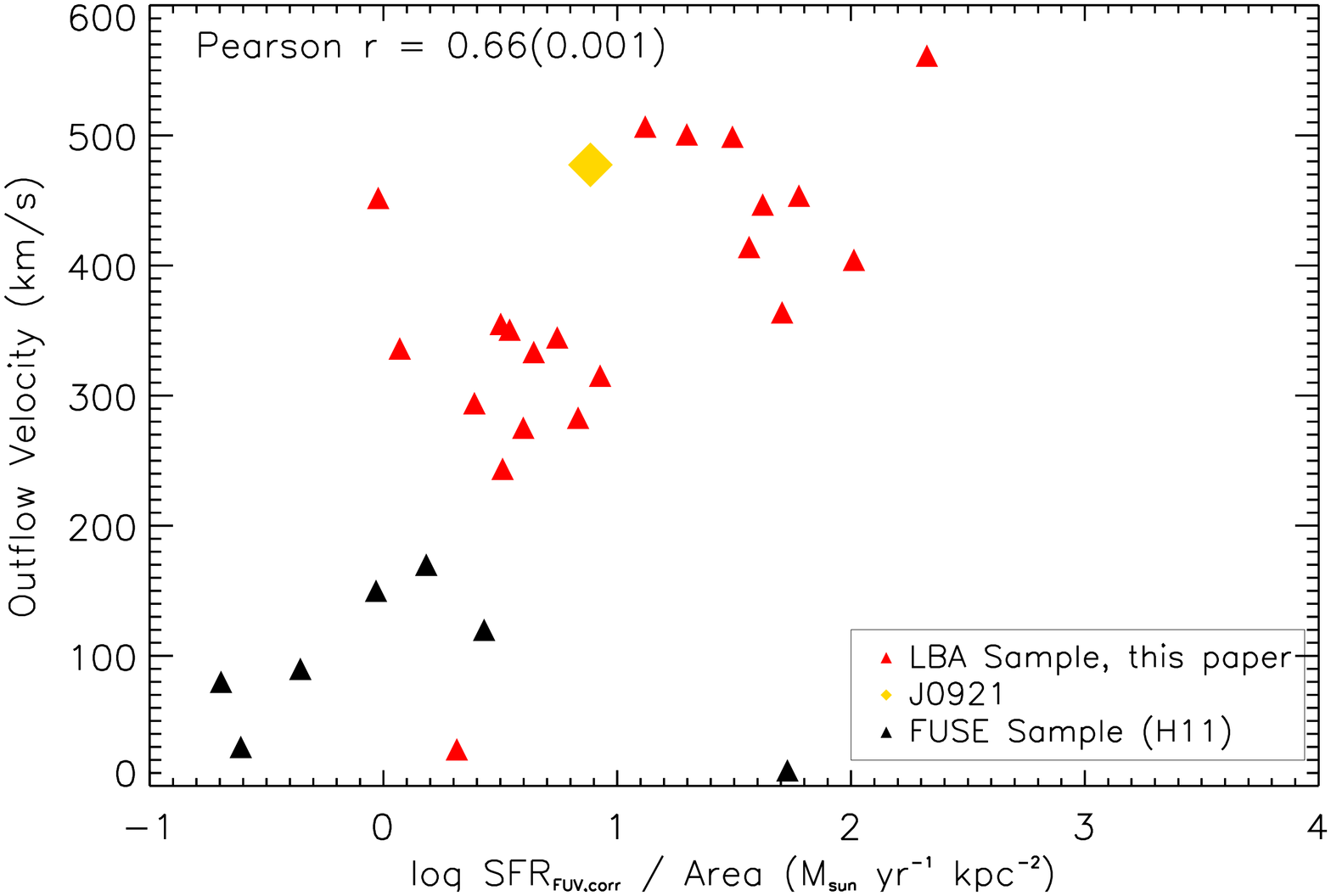}
\caption{\small Correlation between SFR per unit area and outflow velocity.  The SFR is derived from the IR-corrected FUV luminosity to account for dust obscuration and the area is measured from the FUV images taken previously with ACS for our old sample or from the NUV acquisition images for our new sample using SeXtractor.  The outflow velocity is measured from the flux-weighted mean of the \ion{Si}{3}~absorption feature relative to the systematic velocity.  We include, in black, measured values from local galaxies taken with FUSE measured using the \ion{C}{2}~absorption line to calculate the outflow velocity(see H11 for more details).  The top left lists the calculated Pearson r correlation coefficient as well as the statistical deviation from zero (in parentheses).}
\label{fig:vout_sfrarea}
\end{figure}

H11 argued that LBAs with a dominant, compact starburst were the best candidates for high escape fractions of ionizing continuum radiation. They argued that these objects provide extreme feedback (in the form of radiation pressure, wind ram pressure, and a high intensity of ionizing radiation), and this allows them to blow out, ionize or otherwise disrupt the neutral ISM. We have confirmed and quantified that result, finding that our galaxies ranked as the strongest candidates for high escape fractions of ionizing continuum radiation have the highest rates of star formation per unit area though there is a less clear link between those objects officially classified as DCOs and those that are ``leaky".  While all of our galaxies are chosen to have compact star-forming regions, $R_{50}\lesssim1.0$ kpc, it is our most compact sources that make the best candidates for leaking Lyman continuum. We also confirm the result in H11 that the leaky galaxies are characterized by high wind outflow speeds. We show the connection between starburst compactness and outflow speed explicitly in Figure \ref{fig:vout_sfrarea}. Thus, all of these relations create a consistent picture where concentrated massive starbursts are able to drive strong winds capable of creating holes in the optically-thick ISM through which ionizing radiation might escape.

\section{Conclusion}
\label{sec:conclusion}

In this paper we have reported on our analysis of a sample of 22 local ``Lyman Break Analogs'' (LBAs) using HST COS UV-spectroscopy plus ancillary ultraviolet, optical, and infrared data. Our goals were to use these data to better understand how to recognize galaxies that are likely to have escaping Lyman continuum radiation, and to obtain new insights into the conditions and processes that enable ionizing radiation to escape. We have argued that these goals can be best accomplished using relatively nearby galaxies.  We have fit SB99 models of instantaneous and continuous starbursts to characterize the stellar population of each LBA, and find these are young starbursts with the prominent populations of O stars needed to produce substantial amounts of ionizing radiation. We have subtracted the resulting model spectra of the starburst to create a spectrum that shows only features due to the interstellar medium (ISM). By examining the low-ionization ISM lines we see evidence for only partial coverage of the starburst by outflowing neutral gas: while the absorption features are saturated (optically thick), the flux in the line core does not go to zero. This is corroborated by our measurements showing cases in which a significant relative amount of the \lya~emission line profile is blue-shifted relative to the galaxy systemic velocity. This indicates, most likely, that HI only partly covers the front-side of the galactic outflow. Finally, we use SDSS optical spectra to show that the [SII] emission line doublet is unusually weak in many of our galaxies (compared to typical SDSS galaxies). This may indicate the presence of matter-bounded HII regions (gas that is optically thin to the Lyman continuum).

We have found that these three indirect indicators of ``leakiness'' all correlate well with one another. We have therefore combined all three diagnostics to rank-order our galaxies in terms of likely leakiness. We have noted that the inference of leakiness is corroborated by the direct detection of escaping ionizing radiation from the galaxy we rank number 3 out of 22 in our sample (Borthakur et al. 2014). We have also found that our leakiness rank is well-correlated with the equivalent width of the \lya~emission line, which has been widely suggested as a diagnostic of escaping ionizing radiation. Two other proposed diagnostics did not correlate with our leakiness rank: the ratio of the star-formation rates measured using the far-UV vs. \halpha~dust-corrected (intrinsic) luminosities, and the flux ratio of the optical [OIII]5007/[OII]3727 emission lines. Evidently other factors are more important in determining these ratios in our sample of galaxies.  

We then correlated our leakiness rank with the basic properties of the galaxy and its starburst. We found that the strongest correlations of leakiness were with the compactness of the starburst (SFR/area) and with the outflow speed measured using the ISM absorption lines. We also showed that the outflow speed is strongly correlated with the SFR/area. We have therefore argued that the extreme feedback associated with a compact starburst is responsible for creating holes in the neutral ISM that allow ionizing Lyman continuum photons to escape. This is likely to be accomplished through a combination of the high intensity of the ionizing radiation field and the strong outward pressure exerted by the radiation and hot starburst-driven wind. 

A similar, but more ubiquitous population of compact star-forming galaxies at high redshift could create escape fractions high enough to re-ionize the universe.  In addition to providing these new clues to the processes that allow ionizing radiation to escape gas-rich galaxies, our analysis has established several indirect indicators that can be used to identify plausible ``leaky'' galaxies at high-redshift. This is crucial because the opacity of the inter-galactic medium makes it impossible to directly observe the escape of ionizing radiation from galaxies at or near the epoch of re-ionization.  

\acknowledgements
We would like to thank the referee for their comments and suggestions that helped improve this paper.

R.A is supported under a grant for program number 13017 provided by NASA through a grant from the Space Telescope Science Institute, which is operated by the Association of Universities for Research in Astronomy, Incorporated, under NASA contract NAS5-26555

Some of the data presented in this paper were obtained from the Mikulski Archive for Space Telescopes (MAST). STScI is operated by the Association of Universities for Research in Astronomy, Inc., under NASA contract NAS5-26555. Support for MAST for non-HST data is provided by the NASA Office of Space Science via grant NNX13AC07G and by other grants and contracts.

Funding for the SDSS and SDSS-II has been provided by the Alfred P. Sloan Foundation, the Participating Institutions, the National Science Foundation, the U.S. Department of Energy, the National Aeronautics and Space Administration, the Japanese Monbukagakusho, the Max Planck Society, and the Higher Education Funding Council for England. The SDSS Web Site is http://www.sdss.org/.

The SDSS is managed by the Astrophysical Research Consortium for the Participating Institutions. The Participating Institutions are the American Museum of Natural History, Astrophysical Institute Potsdam, University of Basel, University of Cambridge, Case Western Reserve University, University of Chicago, Drexel University, Fermilab, the Institute for Advanced Study, the Japan Participation Group, Johns Hopkins University, the Joint Institute for Nuclear Astrophysics, the Kavli Institute for Particle Astrophysics and Cosmology, the Korean Scientist Group, the Chinese Academy of Sciences (LAMOST), Los Alamos National Laboratory, the Max-Planck-Institute for Astronomy (MPIA), the Max-Planck-Institute for Astrophysics (MPA), New Mexico State University, Ohio State University, University of Pittsburgh, University of Portsmouth, Princeton University, the United States Naval Observatory, and the University of Washington.

\clearpage
\LongTables
\begin{landscape}

\begin{deluxetable}{p{0.4in}|l|l|r|r|r|r|c|r|r|r|r|r|r|r|r|r|r|}
\tabletypesize{\scriptsize}
\tablewidth{0pt}
\tablecolumns{17}
\tablecaption{General LBA Characteristics\label{tab:lba}}
\tablehead{\colhead{Name\tablenotemark{1}}&\colhead{Coordinates\tablenotemark{2}}&\colhead{z\tablenotemark{3}}&\colhead{$R_{50}$\tablenotemark{4}}&\colhead{$M*$\tablenotemark{5}}&\colhead{$M*_{b}$\tablenotemark{6}}&\colhead{DCO\tablenotemark{7}}&\colhead{SFR$_{FUV}$\tablenotemark{8}}&\colhead{EW$_{H\alpha}$\tablenotemark{9}}&\colhead{SFR$_{H\alpha}$\tablenotemark{10}}&\colhead{${{FUV} \over {H\alpha}}$\tablenotemark{11}}&\colhead{$D_{\sii}$\tablenotemark{12}}&\colhead{v\tablenotemark{13}}&\colhead{F$_{res}$\tablenotemark{14}}&\colhead{$R_{eqw}$\tablenotemark{15}}&\colhead{EW\tablenotemark{16}}&\colhead{Rank\tablenotemark{17}}}
\startdata
\sidehead{Earlier Program}
\tableline
J0055&00:55:27.46-00:21:48.7&0.16744&0.32&9.7&8.56&0&23.58&375&28.76&0.82&0.062&414&1.04&-1.25&2.32&20\\
J0150&01:50:28.40+13:08:58.3&0.14668&1.37&10.3&8.15&0&37.44&199&19.96&1.88&0.063&355&6.52&-1.72&3.04&18\\
J0213&02:13:48.53+12:59:51.4&0.21902&0.39&10.5&9.55&1&18.96&31&5.7&3.33&0.204&501&84.85&0.69&9.20&1\\
J0808&08:08:44.26+39:48:52.3&0.09123&0.08&9.8&8.24&1&8.49&77&4.26&1.99&0.368&561&91.2&0.26&15.52&2\\
J0921&09:21:59.38+45:09:12.3&0.23499&0.78&10.8&8.37&1&29.36&72&23.43&1.25&0.118&477&72.37&1.04&4.01&3\\
J0926&09:26:00.40+44:27:36.1&0.18072&0.69&9.1&8.42&0&10.37&577&17.56&0.59&0.074&351&78.01&0.14&36.22&4.5\\
J0938&09:38:13.49+54:28:25.0&0.10208&0.67&9.4&8.3&0&11.18&407&16.09&0.69&0.024&275&7.33&-0.64&1.46&19\\
J2103&21:03:58.74-07:28:02.4&0.13689&0.46&10.9&8.86&1&41.36&105&43.14&0.96&0.299&499&40.7&0.03&25.56&4.5\\
\tableline
\sidehead{New Program}
\tableline
HARO11&00:36:52.70$-$33:33:17.0&0.02060&2.19&10.2&7.49&0&35.39&-&-&-&-&336&40.49&-0.26&7.67&10\\
J0021&00:21:01.02+00:52:48.1&0.09840&0.53&9.3&8.34&0&14.91&514&18.65&0.8&0.083&315&71.15&0.05&25.46&6\\
J0823&08:23:54.95+28:06:21.6&0.04722&0.34&8.6&8.97&0&9.57&504&9.41&1.02&0.045&507&1.64&-&-&21\\
J1025&10:25:48.47+36:22:58.4&0.12650&0.61&9.2&8.1&0&7.56&395&12.15&0.62&0.053&244&23.89&0.02&20.71&14\\
J1112&11:12:44.15+55:03:47.1&0.13163&0.33&10.2&8.56&1&28.67&205&24.78&1.16&0.194&447&20.63&-0.63&7.60&9\\
J1113&11:13:23.99+29:30:39.2&0.17514&1.09&9.6&9.59&0&7.09&24&1.25&5.67&0.051&452&48.12&-0.09&0.85&11\\
J1144&11:44:22.31+40:12:21.2&0.12695&0.76&9.9&9.3&0&8.89&85&7.08&1.26&0.013&294&4.98&-2.89&0.78&22\\
J1414&14:14:54.23+05:40:47.6&0.08190&0.63&8.5&8.9&0&5.14&351&6.34&0.81&0.045&28&0&0.28&1.83&16.5\\
J1416&14:16:12.96+12:23:40.5&0.12316&0.19&10&8.2&1&23.35&183&19.95&1.17&0.183&404&14.59&0.5&1.69&7\\
J1428&14:28:56.41+16:53:39.4&0.18167&0.71&9.6&8.61&1&13.92&249&19.81&0.7&0.02&334&36.65&0.04&19.65&15\\
J1429&14:29:47.03+06:43:34.9&0.17350&0.29&9.4&9.46&1&26.8&850&36.04&0.74&0.083&364&25.64&0.27&32.17&8\\
J1521&15:21:41.52+07:59:21.7&0.09426&0.37&9.5&8.03&1&5.85&145&5.96&0.98&0.022&283&49.04&-1.07&3.96&16.5\\
J1525&15:25:21.84+07:57:20.3&0.07579&0.51&9.4&7.79&0&9.05&126&6.35&1.43&0.068&345&21.56&-0.01&16.57&12.5\\
J1612&161245.59+081701.0&0.14914&0.31&10&8.67&1&36.15&174&32.23&1.12&0.163&454&19.87&-0.41&13.60&12.5\\
\enddata
\tablenotetext{1}{Shortened coordinate used as identifier throughout the paper.}
\tablenotetext{2}{Full SDSS J2000 coordinate. The coordinates for HARO11 are from NED\footnote{http://ned.ipac.caltech.edu/}.}
\tablenotetext{3}{SDSS spectroscopic redshift.}
\tablenotetext{4}{Petrosian radius enclosing 50$\%$ of object flux in the UV (kpc). For additional details see section \ref{ssec:dco}}
\tablenotetext{5}{Mass estimate taken from MPA/JHU Catalog ($M_{\bigodot}$). For more information see section \ref{ssec:mass}.}%
\tablenotetext{6}{Burst Mass estimate from Starburst 99 models ($M_{\bigodot}$). For more information see section \ref{ssec:mass}.}
\tablenotetext{7}{1 indicates this has been determined to be a dominant central object (see section \ref{ssec:dco}) from HST photometry.}
\tablenotetext{8}{Star formation rate ($M_{\bigodot}$ yr$^{-1}$) calculated from the FUV luminosity corrected with the observed far-Infrared luminosity according to \citet{Kennicutt2012}. For more information see section \ref{ssec:halpha}}
\tablenotetext{9}{Rest-frame equivalent width of H$\alpha$ (in emission) taken from the MPA-JHU spectroscopic reanalysis of SDSS.}
\tablenotetext{10}{Star formation rate calculated from extinction-corrected \halpha~flux ($M_{\bigodot}$ yr$^{-1}$) using equation 1 of \citet{Overzier2009} with \halpha~flux taken from SDSS.  For more information see section \ref{ssec:halpha}.}
\tablenotetext{11}{Ratio of Infrared-corrected FUV SFR, to the extinction corrected \halpha~SFR.}
\tablenotetext{12}{Perpendicular distance from the star-forming ridge line measured on a plot of log (\ion{S}{2}/\halpha) vs. $\log(\oiii/H\beta)$(dex).  For additional details refer to section \ref{ssec:bpt}}
\tablenotetext{13}{Outflow velocity calculated as the first moment of the \ion{Si}{3}~absorption line (km s$^{-1}$; see section \ref{ssec:velocity} for additional information).}
\tablenotetext{14}{Residual Flux in the \ion{Si}{2}~absorption line measured at the centroid of the \ion{Si}{2}~1260 $\rm \AA$ line (\%).  For a description of the measurement see section \ref{ssec:res_flux}.}
\tablenotetext{15}{{Lyman} $\alpha$ equivalent width ratio of blueshifted/redshift equivalent width where a negative value is absorption and a positive value is emission.  More details are provided in section \ref{ssec:lya}}
\tablenotetext{16}{{Lyman} $\alpha$ equivalent width (EW) measured over the same region as $R_{eqw}$ but calculated using only emission.  More details can be found in section \ref{ssec:lya}}
\tablenotetext{17}{Rank of objects from 1, most ``leaky" to 22, least ``leaky", based on a combination of the distance from the star-forming ridge, residual flux and \lya~EW ratio.  Details of the calculation can be found in \ref{ssec:rank}.}
\tablecomments{Columns 13-17: All measurements can be assumed to have errors on the order of 10\%-15\% dominated by systematics in the polynomial fit to the continuum emission and subtraction of the SB99 models}
\end{deluxetable}

\clearpage
\end{landscape}

\bibliography{lba.bib}

\label{lastpage}

\end{document}